\documentclass[12pt]{article}

\usepackage{a4wide}
\usepackage{amssymb}
\usepackage{slashed}
\usepackage{verbatim}

\usepackage[intlimits]{amsmath}

 \numberwithin{equation}{section}

\usepackage{amsfonts}

\usepackage{epsfig}

\newcommand{\be}{\begin{equation}}
\newcommand{\ee}{\end{equation}}
\newcommand{\bea}{\begin{eqnarray}}
\newcommand{\eea}{\end{eqnarray}}

\begin{document}

\setcounter{table}{0}

\begin{flushright}\footnotesize

\texttt{ICCUB-19-007}

\end{flushright}

\mbox{}
\vspace{0truecm}
\linespread{1.1}

\vspace{0.5truecm}




\centerline{\Large \bf  Properties of the  partition function of } 

\medskip

\centerline{\Large \bf   ${\cal N}=2$ supersymmetric QCD  with massive matter}

\vspace{1.3truecm}

\centerline{
    {\large \bf J. G. Russo} }

\vspace{0.8cm}

\noindent  
{{\it Instituci\'o Catalana de Recerca i Estudis Avan\c{c}ats (ICREA),\\
Pg. Lluis Companys, 23, 08010 Barcelona, Spain.}}

\medskip
\noindent 
{{\it  Departament de F\' \i sica Cu\' antica i Astrof\'\i sica and Institut de Ci\`encies del Cosmos,\\ 
Universitat de Barcelona, Mart\'i Franqu\`es, 1, 08028
Barcelona, Spain. }}

\medskip

\noindent  {\it E-Mail:}  {\texttt jorge.russo@icrea.cat} 

\vspace{1.2cm}

\centerline{\bf ABSTRACT}
\medskip

We study the different quantum phases that occur in massive ${\cal N}=2$  supersymmetric QCD with gauge groups $SU(2)$ and $SU(3)$ as the coupling $\Lambda/M$ is gradually increased from 0 to infinity.
The phases can be identified by computing the exact partition function by saddle-points, combining supersymmetric localization and the Seiberg-Witten formalism. 
In all cases, we find two phases, a weak coupling and a strong coupling phase,  separated by a  critical point described by a  superconformal field theory or involving superconformal sectors. In crossing the critical point, the dominant saddle-point hops from one singularity of the curve to another one.
The theories seem to undergo a second-order phase transition with divergent susceptibility.

\noindent

\vskip 1.2cm
\newpage

\def\sech{ {\rm sech}}
\def\p{\partial}
\def\pa{\partial}
\def\ov{\over }
\def\a{\alpha }
\def\g{\gamma}
\def\s{\sigma }
\def\td{\tilde }
\def\vp{\varphi}
\def\strokedint{\int}
\def \ha {{1 \over 2}}

\def\KK{{\cal K}}

\newcommand\cev[1]{\overleftarrow{#1}}



\textwidth = 450pt
\hoffset=-30pt

\tableofcontents

\bigskip


\section{Introduction}


Since the works by Seiberg and Witten \cite{Seiberg:1994rs,Seiberg:1994aj}, there have been profound advances in our understanding
of the strong coupling dynamics of  ${\cal N}=2$ supersymmetric gauge theories in four
dimensions. ${\cal N}=2$ theories exhibit extremely interesting physical phenomena such as asymptotic freedom and electric-magnetic duality, which can be described by means of exact formulas that incorporate all perturbative and non-perturbative contributions in closed form.
The presence of quantum phase transitions in theories with massive matter may lead to new insights into the physics of critical phenomena and into the detailed gauge-theory dynamics driving the phase transitions, within a framework with complete analytic control on scaling relations, critical exponents and universality.
Exact results for some supersymmetric observables can be obtained by  localization techniques. It
has been used, in particular, to compute the partition function of general ${\cal N} = 2$ four-dimensional
theories, with any gauge group and matter content \cite{Nekrasov:2002qd,Nekrasov:2003rj,Pestun:2007rz}.

Here we consider ${\cal N}=2$ supersymmetric QCD with $SU(N)$ gauge group and $N_f<2N$ massive hypermultiplets of equal mass $M$,
an asymptotically free theory. 
An important problem is to understand if the theory has a smooth behavior when the coupling $\Lambda/M$ is varied all the way from 0 to infinity, or, on the contrary, it undergoes quantum phase transitions.
The presence of possible phase transitions is manifested by  non-analytic behavior of physical observables,
such as the free energy $F=- \ln Z$, where $Z\equiv Z^{\rm SQCD}_{N_f}$ is the partition
function.
The partition function of the theory on ${\mathbb S}^4$ can be computed by using supersymmetric localization \cite{Pestun:2007rz}, which reduces it to a finite dimensional integral over the Coulomb branch moduli space, $\langle\Phi \rangle ={\rm diag}\big(a_1,...,a_N\big)$, $\sum_{i=1}^N a_i=0$. More precisely, Pestun's construction selects
an integration contour over the real slice (with our conventions for the $a_i$).
 The partition function is given by the formula:
\be\label{NfQCD-partfunc}
Z^{\rm SQCD}_{N_f} =  \int d^{N-1}a\, \,\frac{\prod_{i<j}^{}\left(a_i-a_j\right)^2 H^2(a_i-a_j)}{\prod_{i}^{}H^{N_f}(a_i+M) }
\,\,{\rm e}\,^{(2N-N_f )\ln\Lambda \sum\limits_{i}  a_i ^2}  \left|\mathcal{Z}_{\rm inst}\right|^2, .
\ee
where   
\begin{equation}\label{functionH}
 H(x)\equiv \prod_{n=1}^\infty \left(1+\frac{x^2}{n^2}\right)^n \,{\rm e}\,^{-\frac{x^2}{n}} .
\end{equation}
Here we have set the radius $R$ of the sphere equal to one. It can be restored by 
$M\to MR$, $\Lambda\to \Lambda R$ and $a_i\to a_i R$.
Computing this integral is obviously difficult, but it can be exactly computed in two limits.
One limit is the large $N$ limit discussed in \cite{Russo:2013kea,Russo:2013sba}.
Another limit is the decompactification limit $R\to\infty $ for low rank groups,  discussed in \cite{Russo:2014nka}.

In the large $N$, Veneziano limit, with fixed $N_f/N$, the instanton factor $\left|\mathcal{Z}_{\rm inst}\right|^2$ becomes equal to one
 and the integral is determined by a saddle-point calculation. 
If one further takes the decompactification limit $R\to\infty$ the one-loop factor simplifies due to the simple asymptotic form of  $\ln H\approx -\frac12 x^2\ln x^2$ at $|x|\gg 1$.
This leads to surprising physical consequences:
the theory undergoes a quantum phase transition of third order at $M=2\Lambda$. This is
dictated by the discontinuity of the third derivative of the free energy with respect to the coupling $\Lambda $
and the non-analytic behavior is due to the appearance of massless components of the hypermultiplet.
On the other hand, the 1/2 supersymmetric circular Wilson loop in the fundamental representation can also be computed exactly and it turns out
to be discontinuous in the first derivative
(discontinuities of the Wilson loop in high antisymmetric representations have also been studied \cite{Russo:2018vun}). 
Similar large $N$ phase transitions appeared in  ${\cal N}=2^*$ theory, but in this case with a more complicated phase structure. The large $N$,  ${\cal N}=2^*$  theory has 
 been thoroughly investigated in a number of works, with impressive matches with holography \cite{Buchel:2013id,Bobev:2013cja,Zarembo:2014ooa,Chen-Lin:2017pay,Bobev:2018hbq,Russo:2019lgq}.

For finite $N$, instanton contributions are not suppressed and their contribution is crucial in order to understand the phase structure of the  theory \cite{Russo:2014nka}.
In such a case, the partition function can be more efficiently determined from  Seiberg-Witten theory.
This yields the prepotential ${\cal F}$ as a holomorphic function of the moduli $a_i$.
The partition function is then given by  \cite{Nekrasov:2002qd}
\be
 Z = \int d^{N-1}a \ \prod_{i<j}^N (a_i-a_j)^2\ |Z_0|^2\ ,
\ee
where $Z_0$ is related to the prepotential by the formula
\be
\label{nek}
2\pi i {\cal F} (a_i) = \lim_{\epsilon_{1,2}\to 0} \epsilon_1\epsilon_2 \ln Z_0\ .
\ee
Here $\epsilon_1,\ \epsilon_2$ are the equivariant deformation parameters, which, for ${\mathbb S}^4$,  must be set $\epsilon_1=\epsilon_2=1/R$  \cite{Okuda:2010ke,Pestun:2007rz}.  Thus one finds
\be
\label{sarr}
\lim_{R\gg 1 }   Z^{\rm SQCD} ({\mathbb S}^4) = \lim_{R\gg 1 }\int d^{N-1}a \ e^{-R^2 S(a_i,M)}\ ,
\ee
where  
\be
S\equiv  -{\rm Re}(4\pi i {\cal F})\ ,
\ee
and we have neglected the Vandermonde determinant in the large $R$ limit as it gives a $1/R^2$ contribution to the action. 
Since $R$ is large, the flat-space partition function $Z= \lim_{R\gg 1 }   Z^{\rm SQCD} ({\mathbb S}^4) $ is determined by saddle-points of $S$. 
This was the basic idea in  \cite{Russo:2014nka} and important aspects of this approach were later
made more precise in \cite{Hollowood:2015oma}, in the context of  ${\cal N}=2^*$  theory with gauge group $SU(N)$.
It follows that the exact free energy of the theory is given by 
\begin{equation}
    F(\Lambda/M)=-\ln Z=-R^2\ {\rm Re}(4\pi i {\cal F})\bigg|_{a_i=a_i^*}
    \label{freones}
\end{equation}
where $a_i^*$ are solutions of
\begin{equation}
{\rm Im}    \frac{\partial {\cal F}}{\partial a_i} = {\rm Im} (a_{Di}) =0\ ,\qquad i=1,...,N-1\ ,
\label{adaa}
\end{equation}
and $a_{Di}$ are the usual dual magnetic variables. It should be noted that 
the resulting free energy depends only on the coupling $\Lambda/M$ and it does not depend on any moduli
(since the partition function integrates over the $a_i$).
The imaginary part of the period matrix,
\be
\tau_{ij}= \frac{\partial^2 {\cal F}}{\partial a_i \partial a_j}\ ,
\ee
is positive definite, as it represents the metric in the moduli space. This proves that the saddle-point calculation is applicable.

The imaginary parts of the $a_{Di}$ vanish at degenerate
points of the Riemann surface, i.e. 
when some cycles shrink to zero. 
This implies that the path integral is computed by certain critical points of the prepotential  corresponding to $N-1$ massless BPS dyons, i.e. the path integral is dominated by specific singular curves\footnote{As it will be clear in the next sections,  the converse is not true: not every singularity represents a saddle-point of the partition function integral.}. Saddle points typically lie inside particular domains of marginal stability \cite{Hollowood:2015oma} (see \cite{Brandhuber:1996xk,Bilal:1997st} for related studies in $SU(2)$ SQCD).

Using this approach, in \cite{Russo:2014nka} it was shown that the theory with $SU(2)$ gauge group and $N_f=2$  massive flavors
has a phase transition at $\Lambda= 2M$, whose origin is very similar to the large $N$ phase transitions of \cite{Russo:2013kea}: at a specific coupling, a component of the electrically charged hypermultiplet becomes massless.
 The calculation was carried out by determining the relevant saddle-point for
the strong coupling phase $\Lambda> 2 M$ and noting that there is a singular behavior at $\Lambda=2M$.
The origin of this singular behavior is well understood and  it is related to the Argyres-Douglas phenomenon
\cite{Argyres:1995jj}.
${\cal N}=2$ SQCD with massive matter is known to have fixed points at specific values of the parameters, 
where there is a collision of singularities corresponding to the appearance of mutually non-local massless states \cite{Argyres:1995xn,Eguchi:1996vu}.
These fixed points represent interacting superconformal field theories.

The reason for the appearance of mutually non-local massless states is due to the fact that, when calculating the partition function, we are sitting in a saddle-point, which itself implies the presence of $N-1$ massless dyons; as we move with the coupling $\Lambda/M$ from 0 to infinity, there is a critical coupling where the degeneracy of the curve increases and also a component of the electrically charged hypermultiplet becomes massless. From this perspective, it is now clear that the critical points of the phase transitions must correspond
to fixed points of the type studied in \cite{Argyres:1995xn,Eguchi:1996vu}.

The organization of this paper is as follows.
In section 2 we review the hyperelliptic curves describing ${\cal N}=2$ SQCD with massive matter
and explain the general procedure used in the following sections.
Section 3 is devoted to SQCD with $SU(2)$ gauge group and $N_f=0,1,2,3$ flavors.
After discussing the $N_f=0$ case in section 3.1, in section 3.2 
we first consider the $N_f=2$ case to complete the picture initiated in \cite{Russo:2014nka}
by determining the relevant saddle-point also in the weak coupling phase and
computing the order of the phase transition. Sections 3.3 and 3.4 deal with the cases of $N_f=3$ and $N_f=1$.
In general, although the procedure is similar in all cases,  there is no universal pattern and
the analysis must be made case by case.
In section 4, we study SQCD with $SU(3)$ gauge group  and $N_f$ fundamental hypermultiplets, by considering three examples, $N_f=2,3,4$
(which illustrate the cases $N_f<N$, $N_f=N$ and $N_f>N$).  $SU(3)$ SQCD is described by  hyperelliptic curves containing singularities of  more general type 
and many new features appear. Nonetheless, the different phases will be completely characterized in the sense that the partition function
will be uniquely determined in terms of the prepotential evaluated at the specific singularities that represent the dominant saddle-point for a given coupling.
A discussion on the critical behavior is given in section 4.5. The results
are summarized in section 5.

To avoid an excessive overloading of  labels and letters, in  different cases we  make use of the same letters for similar quantities, although
in each subsection they are defined independently. For example, in the case of the $SU(3)$ gauge group discussed in 
section 4, for even $N_f=0,\ 2,\ 4$, the hyperelliptic curve factorizes as $y^2=Q_+Q_-$. While we use the same symbols $Q_+,\  Q_-$, they are obviously  different for the different cases of  $N_f=0,\ 2,\ 4,$ discussed in subsections 4.1, 4.3 and  4.2.

\section{${\cal N}=2$ supersymmetric QCD and phase transitions}


The curve that describes ${\cal N}=2$ supersymmetric
$SU(N)$ gauge theory coupled to  $N_f$ fundamental hypermultiplets 
with arbitrary masses has been found in \cite{Hanany:1995na,Argyres:1995wt}
(the curve describing  $SU(N)$ pure super Yang Mills was found in \cite{Klemm:1994qs,Argyres:1994xh}). For simplicity,
here we consider the case of equal masses.
The hyperelliptic curve is given by
\be
y^2= C(x)^2 -G(x)\ ,
\label{eguch}
\ee
\be
C(x)
= x^{N}+ \sum_{k=2}^{N} x^{N-k} s_k+q(x)\ ,
\qquad
G(x)=\Lambda^{2N-N_f} (x+M)^{N_f}\ ,
\ee
\be
q(x)=\frac14 \Lambda^{2N-N_f}\sum_{k=0}^{N_f-N}  \binom{N_f}{k}
x^{N_f-N-k}M^k\ ,
\ee
where the term $q(x)$ is absent when $N_f< N$. The superconformal case corresponds to $N_f=2N$, $M=0$.
The $a_n$, $a_{Dm}$ are the  periods of a meromorphic  one-form over a basis of homology one-cycles of the curve,
\be
a_n=\oint_{\beta_n} \lambda\ ,\qquad a_{Dm}=\oint_{\alpha_m} \lambda\ ,
\ee
where
\be
\label{merom}
\lambda =\frac{x}{4\pi i}\ d\ln \frac{C-y}{C+y}\ .
\ee


In general, the condition \eqref{adaa} corresponds to points in moduli space where massless dyons appear. It requires that
$N-1$ cycles shrink to zero. In terms of the curve, 
we must demand that $N-1$ roots of $y^2=0$ are double roots, so that 
 the curve takes the form
\be
\label{pares}
y^2= (x-x_1)(x-x_2) \prod_{i=1}^{N-1} (x-c_i)^2\ .
\ee
This gives $N-1$ conditions, which  fix the $s_k$ moduli parameters, leaving a discrete set of independent solutions $\{ s_k\}$.
In other words, the saddle-point equations \eqref{adaa} correspond to certain zeros of the discriminant that lead to a joining of the branching points in pairs.
This is a necessary, but not sufficient, condition, since the cycles that shrink must specifically  be those which set ${\rm Im}\big( a_{Di}\big)=0$ for $i=1,...,N-1$.
 This includes of course the particular solution  $a_{Di}=0$. But other singularities correspond to setting some $a_i=0$ and they do not represent saddle-points of the partition function integral.

Following this method, the  large $N$ limit of the theory was studied in \cite{Russo:2015vva}, 
generalizing the pure ($N_f=0$) $SU(N)$ super Yang-Mills theory previously studied in 
 \cite{Douglas:1995nw, Ferrari:2001mg}. 
The resulting conditions implied by \eqref{adaa} were solved at large $N$ in \cite{Russo:2015vva} by introducing an eigenvalue density.
It was shown that the form (\ref{pares}) implies an integral equation for the eigenvalue density, which (at large $N$) is identical to the one obtained
directly from the localization matrix integral describing the  one-loop localization partition function 
without instantons. 
In this way one exactly reproduces  all the features of the large $N$  phase transitions found in \cite{Russo:2013kea,Russo:2013sba}, this time from the study of singularities in the Seiberg-Witten curve.

In this paper we are interested in finite $N$. We will consider examples with $N=2,3$ which illustrate the general procedure that can be applied to any gauge group
$SU(N)$. 
Singularities of general type appear when two or more branch points coincide, i.e. when the discriminant of $y^2$, viewed as a polynomial in $x$, vanishes.
The discriminant has the general form
\begin{equation}
\Delta (s_k;\Lambda,M)=\Delta_s^{N_f} \Delta_m\ .
\end{equation}
$\Delta_s$ represents the massless s-quark singularity and the power of $N_f$  reflects
the fact that the quarks transform in the fundamental representation of the global $U(N_f)$ symmetry group of the massive theory.
The reason why this is related to a massless s-quark singularity is that, for large $M\gg \Lambda$, the singularity occurs at $u\sim M^2$; in such a case
the vacuum expectation value of the scalar field is $a\sim \pm M$ and it cancels the bare mass of the hypermultiplet.
However, in the strong coupling regime $\Lambda\gg M$ this singularity is associated with the appearance of a magnetically charged massless state.

In general, the discriminant is a polynomial in the multiple variables $s_k$. Consequently, 
the condition $\Delta=0$ has  multiple solutions.
Moreover, there are many distinct solutions which bring the curve to the form (\ref{pares}) and solve \eqref{adaa}. In principle, the
dominant saddle-point may be identified by the least-action principle. 
Computing the action at the singular points is simple in certain limits. 
Alternatively, the phases can be identified in a solid way by several matching  conditions.

In particular, for $M\gg \Lambda$, the hypermultiplets can be integrated out and what remains is  supersymmetric Yang-Mills theory without matter.
By comparing the respective one-loop $\beta $ functions, one finds that the dynamical scale must be
\be
\label{limiteL}
\Lambda_0 =\Lambda^{1-\frac{N_f}{2N}} M^{\frac{N_f}{2N}}\ .
\ee
Therefore the partition function (hence the saddle-points) of the SQCD theory must coincide with the partition function of pure super Yang-Mills theory (SYM) 
upon taking this low-energy limit, $\Lambda\to 0$, $M\to \infty$, with $\Lambda_0$ fixed.
It will be shown  that this implies that the weak coupling
phase $\Lambda\ll M$ is determined by a zero of $\Delta_m$. 
A straight way to see this is that the solutions to $\Delta_s=0$ move to infinity in the limit $M\to \infty$.
On the other hand, we will argue that the strong coupling phase $\Lambda\gg M$ (which is related to the massless limit)
is determined, in most of the cases, by a zero of $\Delta_s$ 
(for $N>2$, we have $N-1$ conditions, so we will also need to impose $\Delta_m =0$). 
This means that the cycles that shrink at strong and weak coupling must be different: there must be a critical value of $\Lambda $
across which the dominant saddle-point jumps to another singularity.

In addition, we must demand continuity at the critical point: the saddle-points of the subcritical and supercritical phase
must coincide on the critical point. It turns out that in many cases this uniquely determines the relevant saddle-points of the two phases.

Additionally, one can explicitly compare the respective actions of the different competing saddle-points.
This is a simple calculation in two cases:
a) the strong coupling limit $\Lambda\gg M$ and b)  the pure SYM limit, $M\to \infty$,\ 
$\Lambda \to 0$, with fixed $\Lambda_0$.
The reason is that in both cases the theory reduces to an asymptotically free theory without
an additional mass scale.
In this case the prepotential satisfies the equation \cite{Sonnenschein:1995hv,Eguchi:1995jh}
\be
\sum_k a_k \frac{\partial {\cal F}}{\partial a_k}- 2{\cal F} = (2N-N_f) \frac{iu}{2\pi}\ ,
\ee
with $u=-s_2=\frac12 \langle {\rm Tr}\, \Phi^2\rangle $. This generalizes the $SU(2)$, $N_f=0$ case first derived in \cite{Matone:1995rx} and discussions on the massive case can be found in \cite{DHoker:1996yyu,Kubota:1997gb,Flume:2004rp}.
Multiplying by $i$ and taking the imaginary part, we find that, on the saddle points where 
${\rm Im} (a_{Dk})=0$, ${\rm Im} (a_{k})=0$, the free energy is given by the remarkable formula
\be
F\bigg|_{\rm a) \ or\ b)} =-R^2 \ (2N-N_f) {\rm Re}(u)\ .
\label{freeenergy}
\ee
In all examples, the theory will have two phases, $\Lambda>\Lambda_c$ and $\Lambda<\Lambda_c$.
The formula \eqref{freeenergy} can be applied in the two limits mentioned above, a) and b),
to compare the actions of competing saddle-points, when they are present.

Beginning with the solution at $\Lambda\gg M$, as $\Lambda/M$ is decreased, a critical point $\Lambda_c$ will appear
when the degeneracy of the  curve increases due to a branch point joining one of $N-1$ double-degenerate
branch points. In all cases, we found that these enhanced singularities involve  superconformal theories: they 
correspond to   Argyres-Douglas points, where mutually non-local states become massless.
This can be seen from the detailed analysis of degeneracy at the fixed point. The general classification of fixed points for $SU(N)$ SQCD was given in \cite{Eguchi:1996vu}, generalizing
the $SU(2)$ case discussed in \cite{Argyres:1995xn}.

\section{Phase transitions in $SU(2)$ SQCD}

For the $SU(2)$ gauge group it is more convenient to use, instead of \eqref{eguch},
the Seiberg-Witten curves given in \cite{Seiberg:1994aj}, where one of the branch points is taken to infinity.
The $SU(2)$ case will be particularly simple because the possible saddle-points are  obtained
from the zeros of the discriminant in a more straightforward way. 
For higher-rank groups, the condition $\Delta=0$ embodies many types of singularities and most of them are not
related to saddle-points of the partition function integral.
A case-by-case analysis is required in order to solve all the conditions ${\rm Im}(a_{Di})=0$
for $i=1,...,N-1$.

We will now investigate phase transitions that may occur as the coupling $\Lambda/M$ is varied from 0 to $\infty $.
These transitions will imply a non-analytic behavior of the free energy.
For $SU(2)$ we have only periods $a, \ a_D$. Once the condition ${\rm Im}(a_{D})=0$ is solved for some $u=u^*(\Lambda/M)$,
the period $a^*=a(u^*)$ is determined and the free energy
 as a function of $\Lambda/M$ is given by  the formula \eqref{freones}.
A sign of a phase transition appears when, upon crossing  a critical coupling $\Lambda_{\rm c}$,
the path integral gets dominated by different saddle-point.
A more direct sign is when the  free energy presents  a non-analytic behavior   around the critical point.
In all cases, the critical point of the phase transition will occur when the s-quark singularity collides with one of the massless monopole
or massless dyon singularities.   The BPS mass formula is 
\be
M_{\rm BPS}=\sqrt{2}|Z|\ ,\qquad Z=  n_m a_D(u)+n_e a(u)+\sum_{i=1}^{N_f} \sigma_i \frac{M_i}{\sqrt{2}}\ ,
\ee
where $\sigma_i$ are integer and half-integer abelian charges.
For  massless dyons, this implies a linear relation between $a_D$ and $a$, with a possible additive mass term, which does not affect the saddle point equation \eqref{adaa} since in our examples the mass is real. 
Our aim is to completely characterize the different phases in each example by identifying the relevant singularity
and to elucidate the main physical features of the phase transition. This procedure implicitly determines the partition function as a function of $\Lambda/M$, through the above formula (\ref{freones}), in a unique way, although deriving an explicit expression in closed form is complicated.
The reason is that the prepotential for SQCD with flavors cannot be expressed in closed form in  terms of standard special functions in the massive case. Some useful discussions can be found in \cite{Ohta:1996fr,D'Hoker:1996nv,Masuda:1996xj,Masuda:1996np,Bilal:1997st,AlvarezGaume:1997fg}.

\subsection{$SU(2)$ gauge group with $N_f=0$}

It is useful to begin with the  $N_f=0$ theory, which describes the low-energy (infinite mass) limit of the $N_f>0$ theories.
As explained in \cite{Seiberg:1994aj}, in order to study this theory as the low energy limit of SQCD theories with massive hypermultiplets,
instead of using the original curve of \cite{Seiberg:1994rs}, it is more convenient to describe the dynamics in terms of the curve
\be
y^2 =x^2(x-u)+\frac14 \Lambda_0^4 \ x = x(x-x_+)(x-x_-)\ ,
\label{saverio}
\ee
where
$$
x_\pm = \frac12 \left(u \pm \sqrt{u^2 - \Lambda_0^4}\right)\ .
$$
With these conventions, particles still have integer charges $n_m,\ n_e$ and  the structure $Z=a_D n_m+a n_e$
is preserved.

The period $a$ is defined by integrating the meromorphic form along a loop that goes around $0$ and $x_-$,
while $a_D$ is defined in terms of a loop around $0$ and $x_+$.
Singularities arise at zeros of the discriminant 
\be
\label{sincero}
\Delta \propto \Lambda_0^8 (u^2-\Lambda_0^4)\ ,
\ee
which gives $u=\pm \Lambda^2_0$. 
Near these points, the curve takes the form 
$$
y^2 = x \left(x\mp \frac 12\Lambda_0^2\right)^2\ .
$$
This gives $a_D=a$ and ${\rm Im}(a_D)=0$. Therefore the singularities at $u=\pm\Lambda_0^2$
represent  saddle-points in the partition function integral. The dominant saddle-point
is $u=\Lambda_0^2$ because the free energy \eqref{freeenergy} is smallest in this case:
\be
F = - 4R^2\Lambda_0^2\ .
\ee
In discussing massive theories with $N_f>0$, the weak coupling phase $\Lambda\ll M$ must be dominated by this saddle-point, so that the low energy dynamics is the same
as in pure SYM. In particular, this ensures a consistent renormalization group flow
from the $N_f>0$ theories to the $N_f=0$ theory upon the identification
\eqref{limiteL}.

\subsection{$SU(2)$ gauge group with $N_f=2$}

This case, with two flavors of equal masses, was discussed in \cite{Russo:2014nka}, where the saddle-point of the strong coupling phase was identified and it was argued that the theory has a phase transition at $\Lambda=2M$. Here we will clarify what happens in the weak coupling phase $\Lambda<2M$
and in addition compute the order of the phase transition.

For two massive flavors, the curve  is \cite{Seiberg:1994aj}
\be
\label{masw}
y^2 = 
\left(x^2- \frac{1} {64} \Lambda^4\right)  (x-u)+ \frac{1}{4} M^2\Lambda^2 \ x - \frac{1}{32}M^2\Lambda^4\ .
\ee
It is convenient to shift $x\to x+u/3$ and write
\be
y^2 = (x-e_1 ) (x-e_2 ) (x-e_3 )  \ ,
\ee
with
\bea
e_{1,3} &=& \frac{u}{6} -\frac{\Lambda^2}{16}\pm \frac{1}{2}\sqrt{ u+\frac{\Lambda^2}{8}+\Lambda M}\sqrt{ u+\frac{\Lambda^2}{8}-\Lambda M}\ ,
\nonumber\\
e_2 &=& -\frac{u}{3} +\frac{\Lambda^2}{8}\ .
\eea
The prepotential ${\cal F}(a)$ is obtained from the formula
\be
a_D = \frac{\partial {\cal F} }{\partial a}\ ,
\ee
where $a$ and $a_D$ are  period integrals of the meromorphic one-form
\be
\lambda = -\frac{\sqrt{2}}{4\pi } \frac{ y\, dx}{x^2-\frac{\Lambda^4}{64}}\ .
\ee
As usual, this determines $a$ and $a_D$ in terms $u$ and, consequently, ${\cal F}(u)$ or ${\cal F}(a)$.
The period $a_D$ corresponds to the integral over the cycle encircling  $e_1$ and $e_2$,
while $a$ is an integral over a  cycle encircling  $e_2$ and $e_3$.
This last cycle  also surrounds  the pole of the  meromorphic one-form $\lambda$ with residue  $M/\sqrt{2}$.
Useful formulas for the periods in terms of elliptic functions can be found in \cite{Masuda:1996xj,AlvarezGaume:1997fg,Bilal:1997st}.

Singularities occur when two branch points collide, i.e. at the zeros of the discriminant
\be
\Delta = \frac{1}{2^{16}} \Lambda ^4 \Delta_s^2 \Delta_{m+}\Delta_{m-}\ ,
\ee
\be
\Delta_s=  \Lambda ^2+8M^2-8u\ ,\quad \Delta_{m\pm}=  \Lambda
   ^2+8u \pm 8 \Lambda  M\ .
\ee
One singularity corresponds to $\Delta_s=0$, 
\be
u_s=M^2+\frac{\Lambda^2}{8}\ .
\ee
The other two singularities, solving  $\Delta_{m\mp}=0$, appear at
\be\label{singo}
u_{1,2}=\pm M\Lambda -\frac{\Lambda^2}{8}\  . 
\ee
At $u=u_s$, the branch point $e_1$ coincides with $ e_2$ {\it provided} $\Lambda >2M$.
In this case one would find $a_D=0$, therefore a solution of the saddle-point equation.
However, when $\Lambda <2M$, at $u=u_s$ one has $e_2=e_3$.\footnote{This has also consequences for
 the stability domains of the BPS spectra  in the $\Lambda <2M $ and $\Lambda>2M$ regimes \cite{Bilal:1997st}.} 
 The singularity at $u=u_s$ no longer represents
a saddle-point, as it does not solve the condition ${\rm Im}\big( a_D\big)=0$. This can be seen by explicit calculation
(see fig. 2a in \cite{Russo:2014nka}).
The saddle-point must hop to another singularity, implying a phase transition.

At the critical coupling
\be
\Lambda_c =2M\ ,
\ee
the three branch points $e_1$, $e_2$, $e_3$ coincide. The curve has a cusp of the form 
$$
y^2=\left(x-\frac12 M^2\right)^3\ .
$$
This  point describes a superconformal field theory of the Argyres-Douglas type  \cite{Argyres:1995jj},
studied in detail in  \cite{Argyres:1995xn}.
As shown below, this fixed point represents nothing but the critical point of a second-order phase transition.

\subsubsection*{Strong coupling phase $\Lambda >2M$}

The saddle-point equation is
\be
\frac{\partial S(a,M)}{\partial a}=0\ \longrightarrow  \ {\rm Im}\left(\frac{\partial {\cal F}}{\partial a} \right)= {\rm Im}(a_D) =  0\ .
\ee
A particular solution is $a_D=0$, which  requires that $e_1\to e_2$. 
As pointed out above, this occurs at  
\be
u_s= M^2+\frac{1}{8}\Lambda^2 \ ,\qquad \Lambda >2M\ ,
\ee
where the curve takes the form
\be
y^2= \left( x-\frac{\Lambda^2}{8}\right)^2 \left( x-M^2+\frac{\Lambda^2}{8}\right)\ .
\ee
In the vicinity of the singularity one has the behavior
\be
a_D\approx c(u-u_s)\ ,\qquad a\approx a_0+\frac{i}{\pi}c_0 (u-u_s)\ln (u-u_s)\ .
\ee
The monodromy around the singularity gives $a_D\to a_D$, $a\to a-2a_D$ and of course it does not affect the existence of a saddle-point.
As the fixed point $\Lambda=2M$ is reached from above, one finds that $e_2\to e_3$, $y^2= \left( x-\frac12 M^2 \right)^3$ and the only contribution to $a$ comes from the mass residue, leading to
\be
a^*= a(u_s) \longrightarrow \frac{M}{\sqrt{2}}\ \qquad {\rm as}\ \ \Lambda\to 2M\ .
\ee
At this critical point, one component of the hypermultiplet becomes massless, because the bare mass $M$  is canceled by the vacuum value, which
in the present conventions is $\sqrt{2}a=M$. The appearance of this massless state at a specific coupling leads to a non-analytic behavior of the free energy.
This is exactly  the same physical origin as the large $N$ phase transitions of \cite{Russo:2013kea}. 

As long as $\Lambda >2M$, the free energy will be given by
$F(\Lambda/M)= -R^2 {\rm Re} \big(4\pi i {\cal F}(a^*)\big)$, where $a^*$ is the value of $a$ at $u=u_s$.
The free energy is thus completely determined in the strong coupling phase
$\Lambda>2M$ in terms of the prepotential as a function of $\Lambda/M$, computed by sitting
on the $u=u_s$ singularity.

\subsubsection*{Weak coupling phase $\Lambda <2M$}

As explained, when $\Lambda<2M$,  $u=u_s$ does not correspond to a saddle-point of the integral.
In this phase, $\Lambda <2M$, the saddle-point equation ${\rm Im}(a_D) =  0$  is satisfied at the singular point
\be
u\to u_1 = M\Lambda -\frac{\Lambda^2}{8}\ .
\ee
At $u=u_1$, $e_1\to e_3$ and the curve becomes
\be
y^2= \left( x-\frac12 M\Lambda +\frac{\Lambda^2}{8}\right)^2 \left( x-\frac{\Lambda^2}{8}\right)\ .
\ee
$u=u_1$ describes the dyon singularity
\be
a-\frac{M}{\sqrt{2}}-a_D =0\ .
\ee
We have checked that ${\rm Im}(a_D)=0$ using the explicit formulas for $a_D,\ a$ in terms of elliptic functions given in \cite{Bilal:1997st}.

Substituting $u\to u_1$ into the action, one can determine the free energy in the weak coupling phase $\Lambda <2M$.

A  check that the weak coupling phase has been correctly identified is that the corresponding
partition function must match the partition function of pure SYM without matter in the $M\to\infty $
limit where the hypermultiplet is decoupled.
In the current $N_f=2$ case, according to (\ref{limiteL}), the limit corresponds to $M\to\infty$,
$\Lambda\to 0$, with $M\Lambda\equiv \Lambda_0^2=$ fixed. In this limit, the $N_f=2$ curve reduces to the Seiberg-Witten curve \eqref{saverio}
and the singularity at $u=u_1$ approaches the singularity $u=+\Lambda_0^2$ of \eqref{sincero}.
The remaining singularity at $u_2$ is  ruled out by the matching conditions:  it does not match $u_s$ at the critical point where the two phases meet, nor it matches the dominant saddle-point $u= \Lambda_0^2$ of the pure SYM limit.

\subsubsection*{Free energy and critical behavior}

In the subcritical phase $\Lambda< 2M$ the only possible saddle-point occurs at $u=u_1$.
However, in the strong coupling phase $\Lambda< 2M$ there are two competing saddle-points, $u_s$
and $u_1$. In this case both values of $u$ solve the condition ${\rm Im}(a_D)=0$.
The dominant saddle-point is the one with least action. Taking the limit $\Lambda/M\to \infty$
or, equivalently, the massless limit $M\to 0$, we can make use of the formula \eqref{freeenergy}
to obtain $F=-2R^2 u$. As in this limit $u_s\to \Lambda^2/8$ and $u_1\to -\Lambda^2/8$,
the dominant saddle-point of the strong coupling phase is therefore at $u=u_s$, where
\be
F \bigg|_{\Lambda\gg M}= - \frac14 R^2\Lambda^2\ .
\ee
Thus we have a phase transition where, in crossing  $\Lambda_c =2M$, the saddle-point hops from one
singularity to a different one. The order of the phase transition can be computed by
looking at the critical behavior of the free energy.
We begin with the  prepotential, which can be computed as
\be
{\cal F}(u)-{\cal F}(u_0) = \int_{u_0}^{u} du\ a_D(u) \partial_u a(u) \ ,
\ee
where $u_0$ is any generic point on the real line.
We are interested in non-analytic behavior of the free energy at the critical point.
The integrals can be carried out explicitly in the neighborhoods of the conformal point. 
The behavior of $a$ and $a_D$ near the critical point is determined by the scaling dimension
of the perturbation associated with $u$, computed in \cite{Argyres:1995xn}.
For $N_f$ flavors, one has 
\be
\label{uudim}
[u]=\frac{12}{11-N_f}\ .
\ee
As usual, we impose that the periods have dimension 1. For the $N_f=2$ theory, this gives
\be
a-\frac{M}{\sqrt{2}} \sim (u-u_c)^{\frac34}\ ,
\quad  a_D\sim (u-u_c)^{\frac34}\ ,
\qquad u_c=\frac32 M^2\ .
\ee
This behavior is consistent with the explicit expressions in terms of elliptic functions given in \cite{Masuda:1996np,Bilal:1997st}.
Then
\be
{\cal F}(u)\sim {\rm const.}\left(u-u_c \right)^{\frac32} +{\rm analytic}\ .
\ee
Now, particularizing for $u=u_s$ when $\Lambda \geq 2M$, we see that
$u_s-u_c\approx {\rm const.} (\Lambda-2M) $, so ${\cal F}\sim  (\Lambda-2M)^{\frac32}$.
For $\Lambda < 2M$, one also has $u_1-u_c\approx {\rm const.} (\Lambda-2M) $ so ${\cal F}$ has the same critical exponent as $\Lambda $ approaches $2M$ from below.
As a result, the second derivative with respect to $\Lambda $, representing the susceptibility $\chi $,   diverges,
\be
\chi = -\frac{\partial^2F}{\partial\Lambda^2}\sim \frac{1}{\sqrt{\Lambda-2M}}\ ,
\ee
with a critical exponent equal to $-1/2$.
Therefore the phase transition is of the second  order. 

This may be compared with the 
analogous phase transition occurring in the large $N$ SQCD model, \cite{Russo:2013kea,Russo:2013sba}, which is third order.
In conclusion,  the SQCD $SU(2)$ theory with two flavors has
a  phase transition of a similar origin as the large $N$ phase transition found in
SQCD with $N_f<2N$ flavors discussed in \cite{Russo:2013kea,Russo:2013sba}. Just as in the large $N$ phase transitions, the discontinuous behavior is produced by the contribution of massless hypermultiplets
to the free energy at the critical point.
Unlike the large $N$ phase transitions of \cite{Russo:2013kea,Russo:2013sba}, where instantons are suppressed and played no role, here the phase transition
is induced by instantons \cite{Russo:2014nka}.
The superconformal fixed point  represents the critical point of this transition.

\subsection{$SU(2)$ gauge group with $N_f=3$}

We consider three flavors of equal masses $M$. 
The Seiberg-Witten curve is described by 
\be
\label{maswww}
y^2 =x^2(x-u)-\frac{1} {64} \Lambda^2 \left(x- u \right)^2  - \frac{3}{64} M^2\Lambda^2 \ (x-u) +\frac{1}{4}M^3\Lambda x- \frac{3}{64}M^4\Lambda^2\ .
\ee
It has singularities at the zeros of the discriminant
\begin{equation}
    \Delta=\Delta_s^3 \Delta_+\Delta_- =c\Lambda^2 \, (u-u_s)^3(u-u_1)(u-u_2)\ ,
\end{equation}
where  $c$ is an unimportant numerical constant and
\begin{equation}
    u_s=M^2+\frac18 M\Lambda\ ,\qquad  u_{1,2}=\frac{1}{512}\left( \Lambda^2 -96M\Lambda \pm \sqrt{\Lambda(\Lambda+64 M)^3} \right)\ .
\end{equation}
Let us write the curve in the form
\be
y^2=(x-x_0)(x-x_+)(x-x_-)\ .
\ee
In the classical region $u\gg M^2,\Lambda^2 $, the branch points behave as follows
\be
x_0\approx u\ ,\qquad x_\pm \approx \pm \frac{i\Lambda}{8 }\ \sqrt{u}\ .
\ee
We define $a$  in terms of a cycle surrounding $x_-$ and $x_+$, and $a_D$ in terms of a cycle surrounding
$x_0$ and $x_-$. 

We  now consider the behavior of branch points $x_0,\ x_+,\ x_-$ at the three different singularities.
Sitting on the s-quark singularity $u=u_s$ gives $x_+ = x_-$ for any real (positive) value of $\Lambda/M$.  At this singularity we have
\begin{equation}
    y^2\Big|_{u=u_s}=\left( x- \frac{1}{8}M\Lambda \right)^2
    \left(x-M^2+\frac{1}{8}M\Lambda-\frac{\Lambda^2}{64}\right)\ .
\end{equation}
This, however, gives $a=M/\sqrt{2}$, leading to a massless electric hypermultiplet, 
but it does not constitute a solution of the saddle-point equations because ${\rm Im}\big(a_D\big)\neq 0$.
Near this point, one has the behavior
\be
a-\frac{M}{\sqrt{2}}\approx c(u-u_s)\ ,\qquad a_D\approx a_{D0}-\frac{3i}{2\pi} c(u-u_s)\ln (u-u_s)\ , 
\ee
where $a_{D0}$ depends on $\Lambda/M$. This leads to the monodromy
$a\to a$, $a_D\to a_D+3a-3M/\sqrt{2}$.
Similarly, the singularity at $u=u_2$ also sets $x_+=x_-$ and is not a solution of the saddle-point equation.

On the other hand, sitting on the monopole singularity $u=u_1$, the elliptic curve takes the form
$ y^2=(x- x_+)(x- x_0)^2 $; the branch point $x_0$ coincides with $x_-$ for any real value of $\Lambda/M >0$ . As a result, $a_D=0$ (with a monodromy similar as in the $N_f=2$ case). 
This represents the saddle-point of the partition function integral for any coupling.
In particular, note that there is a smooth match with the pure SQCD limit $M\to \infty$ with $\Lambda_0^4=\Lambda M^3$ fixed. In this limit, one has
$ u_1\to \Lambda_0^2$. 
When $\Lambda/M$ becomes equal to 8, the curve has a cusp singularity, $y^2= (x-M^2)^3$. This represents
a superconformal theory studied in \cite{Argyres:1995xn}.

At the critical point,  $u_s=u_1$, there are mutually non-local states becoming massless, the squark and the monopole.
Indeed, the critical point may  be directly obtained by substituting
$u=u_s$ in $\Delta_+\Delta_- $ and demanding that $\Delta_+\Delta_-=0$.
This gives
\begin{equation}
    \Delta_+\Delta_- \Big|_{u=u_s} ={\rm const.}\ (\Lambda-8M)^3\ ,
\end{equation}
which vanishes precisely at the critical point, where there is higher degeneracy.

The free energy  is computed as in previous cases by
evaluating the prepotential at the singularity, this time at $u=u_1$, for any $\Lambda/M$,
\begin{equation}
    F=-\ln Z = -R^2 {\rm Re} \big( 4\pi i{\cal F}\big)\bigg|_{u=u_1}\ .
\end{equation}
In particular, in the strong coupling limit, we can use \eqref{freeenergy} to obtain
the formula
\be
F \bigg|_{\Lambda\gg M}=-R^2 u_1\bigg|_{\Lambda\gg M}= - \frac{R^2\Lambda^2}{256}\ .
\ee
Note that this value of $F$ is smaller than the one obtained by sitting on $u_s$ (which goes to 0 at $M\to 0$). This also reassures that we are in the correct saddle-point.
%
%
%
%
%
While there is no jumping of saddle-points dominating the path integral in going across  the critical point,  the free energy  still presents a non-analytic behavior due to the existence of a fixed point
with superconformal symmetry. From the scaling dimension \eqref{uudim} of the perturbation associated with $u$,
in neighborhoods of the fixed point one obtains the  behavior
\be
a\sim (u-u_c)^{\frac23}\ ,\quad  a_D\sim (u-u_c)^{\frac23}\ ,\qquad u_c= 2M^2\ .
\ee
This gives
\be
 {\cal F}\sim {\rm const.} (u-u_c)^{\frac43}\ .
\ee
The dependence on $\Lambda $ is obtained by sitting on $u=u_1$
and using $u_1-u_c\sim \frac{M}{8}(\Lambda-8M)$.
Like in the two-flavor case, the susceptibility, $\chi =-\partial_\Lambda^2 F$, is divergent, indicating a second-order phase transition.





\subsection{$SU(2)$ gauge group with $N_f=1$}

The Seiberg-Witten curve is 
\be
\label{maszzz}
y^2 = x^2  (x-u)+ \frac{1}{4} M\Lambda^3 \ x - \frac{1}{64}\Lambda^6=(x-x_0)(x-x_+)(x-x_-)\ .
\ee
The discriminant is now
\begin{equation}
    \Delta= -\frac{\Lambda^6}{16}\left( u^3-M^2u^2-\frac98M\Lambda^3 u+\Lambda^3 M^3+\frac{27\Lambda^6}{256}\right)\ .
    \label{disqi}
\end{equation}
Thus singularities arise at three roots $u_s,\ u_1,\ u_2$ of the cubic polynomial equation $\Delta=0$. They are distinguished by the behavior at $M\to \infty $,
where 
\be
u_s\approx M^2\ ,\qquad u_{1,2}\approx \pm \Lambda_0^2\ ,\quad \Lambda_0^4\equiv M\Lambda^3\ .
\ee
The critical point occurs when two of these roots meet. That is, when the discriminant of
$\Delta$ vanishes. This gives
\begin{equation}
    \Lambda = e^{\frac{2\pi i \ell}{3}} \frac{4M}{3},\ \qquad \ell =0,\ 1,\ 2\ .
\end{equation}
At any of these points, there are mutually non-local states becoming massless (quarks, monopoles or dyons).
The corresponding interacting superconformal field theory is  discussed in \cite{Argyres:1995xn}.
As we are examining the behavior of the theory for real $\Lambda $ between 0 and infinity, the only relevant fixed point will be    $\Lambda_c =  \frac{4M}{3}$.
At the fixed point, one has $u_1=u_s$ and $y^2=(x-4M^2/9)^3$. 

In the classical limit $u\gg M^2,\ \Lambda^2 $, the branch points behave as follows
\be
x_0\approx u+O\big(u^ {-1}\big)\ ,\qquad x_\pm \approx \pm \frac{i}{8} \Lambda^3 \frac{1}{\sqrt{u}}+O\big(u^ {-1}\big)\ .
\ee
$a$ is defined as the period integral over the cycle looping around $x_-$ and $x_+$, while $a_D$ as the period integral over the cycle looping around $x_-$ and $x_0$.

One finds the following behavior at the singularities:
\bea
&& u=u_s:\qquad x_0=x_-,\ \ {\rm for}\ \ \Lambda>\frac{4M}{3} ; \qquad x_+=x_-,\ \ {\rm for}\ \ \Lambda<\frac{4M}{3}\ ;
\nonumber\\
\nonumber\\
&& u=u_1:\qquad x_0=x_+,\ \ {\rm for}\ \ \Lambda>\frac{4M}{3} ; \qquad x_0=x_+,\ \ {\rm for}\ \ \Lambda<\frac{4M}{3}\ ;
\nonumber\\
\nonumber\\
&& u=u_2:\qquad x_+=x_-,\ \ {\rm for}\ \ \Lambda>\frac{4M}{3} ; \qquad x_+=x_-,\ \ {\rm for}\ \ \Lambda<\frac{4M}{3}\ .
\nonumber
\eea
In the weak coupling phase $\Lambda<4M/3$,  $u_s,\ u_1,\ u_2$ are real and all three  roots $x_0,\ x_+,\ x_-$ are real.
As a result, the dyon singularity at $u=u_1$, where $x_0=x_+$,  gives ${\rm Im}\big( a_D\big)=0$ with $a_D=a$ (modulo real mass terms).
Therefore $u=u_1$ represents the saddle-point of the weak coupling phase.
As a consistency check, we find that $u=u_1$ matches the saddle-point of the pure ($N_f=0$) SQCD theory in the limit $M\to\infty $ with $\Lambda_0^4\equiv M\Lambda^3 $ fixed, i.e. $u_1\to \Lambda_0^2$. This is not the case for the singularity at $u=u_s$, which moves to infinity in this limit.

Consider now the strong coupling phase $\Lambda >4M/3$. 
In this phase, $u_2$ remains real while $u_s$ and $u_1$ become complex conjugate, $u_s=u_1^*$
(at the critical point, one has $u_s=u_1$).
$u=u_s$ provides a solution of the saddle-point equation, with $a_D=0$, corresponding to a monopole singularity. The saddle-point at $u=u_s$ is unaffected under monodromy, because near a monopole singularity $a_D$ is invariant.
At $u=u_1$, one has  $a_D=a$, corresponding to a dyon singularity.
The fact that $u_1$ and $u_s$ are complex conjugate suggests that both singular points 
equally contribute to the partition function integral. 
We can check the consistency of this picture by repeating the analysis, but now using the curve \eqref{eguch}.
For the present case, it is given by
\be
y^2= \left(x^2-u\right)^2 -\Lambda^3 (x+M)\ .
\ee
The discriminant is still given by \eqref{disqi} (modulo an overall numerical constant). Sitting on $u=u_s$, at the critical point
$\Lambda_c =  \frac{4M}{3}$,
the curve has the cusp singularity
\be
y^2 =\left(x+\frac{2M}{3}\right)^3 \left(x- 2M\right)\ .
\ee
At large $u$, the four branch points have the behavior
\be
x_{1,2}\approx -\sqrt{u} \mp \frac{i\Lambda^{\frac32}}{2u^{\frac14}}\ ,\qquad
x_{3,4}\approx \sqrt{u} \mp \frac{\Lambda^{\frac32}}{2u^{\frac14}}\ .
\ee
We define the periods $a,\ a_D$  in terms of one-cycles surrounding $\{ x_1,\ x_2\}$ and $\{ x_2,\ x_3\}$, respectively.
Then we find that, at $u=u_1$, $x_2=x_3$ for all $\Lambda/M$. Hence $a_D=0$ on the $u=u_1$ singularity.
On the other hand, at $u=u_s$, we have $x_2=x_3$ for $\Lambda>\Lambda_c$ but  $x_1=x_2$ for $\Lambda<\Lambda_c$.
Thus, $u=u_s$ solves the saddle-point equation ${\rm Im}\big(a_D\big)=0$ only in the strong coupling phase. In this phase
the branch points with $u=u_s$ are related to the branch points with $u=u_1$ by complex conjugation.


In the strong coupling limit, $\Lambda\gg M $, one has
\be
u_s\bigg|_{\Lambda\gg M} \approx \frac{3}{2^{\frac{11}{3}}} \big(1-i\sqrt{3}\big)\Lambda^2\ ,
\qquad u_1\bigg|_{\Lambda\gg M} \approx \frac{3}{2^{\frac{11}{3}}} \big(1+i\sqrt{3}\big)\Lambda^2\ .
\ee
Using \eqref{freeenergy}, we find that, either at $u=u_1$ or $u=u_s$, the free energy approaches the value
\be
F\bigg|_{\Lambda\gg M}=-\frac{9}{2^{\frac{11}{3}}}\ R^2\Lambda^2\ .
\ee
In conclusion, the weak coupling phase $\Lambda<4M/3$  is characterized by sitting on $u=u_1$ whereas in  the strong coupling phase $\Lambda>4M/3$ there  are  two saddle-points at $u=u_s$ and $u=u_1$, with $u_s=u_1^*$ and identical contribution to the partition function integral.


Let us now consider the critical behavior. For the theory with $N_f=1$ flavors, from \eqref{uudim} one has 
\be
a\sim (u-u_c)^{\frac56}\ ,\quad  a_D\sim (u-u_c)^{\frac56}\ ,\qquad u_c= \frac{4M^2}{3}\ .
\ee
Near the superconformal point, the prepotential is therefore of the form
\be
{\cal F}\sim {\rm const.} (u-u_c)^{\frac{5}{3}}\ .
\ee 
This gives again a divergent susceptibility $\chi =-\partial_\Lambda^2 F$, indicating that the theory undergoes
a second-order phase transition.


\section{SQCD with gauge group $SU(3)$}

For higher rank groups, the structure of singularities in the Coulomb branch is more complicated.
The theory has different types of singularities (not only double degeneracy or cusps).
Accordingly, the phase dynamics becomes more intricate;
now we need to solve all the conditions ${\rm Im}\left(a_{Di}\right)=0$ and follow the motion of all branch points
along the renormalization group flow. 
Here we will fully characterize the phases and implicitly determine the partition functions of the various
theories by identifying the dominant saddle-points and the specific singularities that govern the critical points of the phase transitions.
These singularities represent
superconformal points that have been classified in  \cite{Eguchi:1996vu}.  Some  aspects of the structure of the theories at these singular points
have been more recently elucidated in \cite{Gaiotto:2010jf,Giacomelli:2012ea,Bolognesi:2015wta}.

\subsection{$SU(3)$ gauge group with $N_f=0$}

It is useful to review some features of the pure super Yang-Mills theory with gauge group $SU(3)$ without matter multiplets,
as this theory describes the low-energy limit of the theories with massive flavors considered in the subsections that follows.
In this case the curve can be written as \cite{Klemm:1994qs,Argyres:1994xh}
\be
y^2=Q_+(x) Q_-(x)\ ,\qquad Q_\pm(x)\equiv  (x^3-u x-v)\pm \Lambda_0^3\ .
\label{nfcero}
\ee
The fixed points of this theory have been studied in  \cite{Argyres:1995jj}. 
Because $Q_+-Q_-=2\Lambda_0^3$, $Q_+$ and $Q_-$ cannot have the same zeros, as we assume $\Lambda_0\neq 0$.
Therefore, the only possibility for four branch points joining pairwise is by demanding that the discriminants $\Delta_\pm$ of both polynomials
$Q_+$ and $Q_-$ are simultaneously equal to zero.
One has
\be
\Delta_\pm = 4u^3-27 (v\mp \Lambda_0^3)^2=0\  \ \longrightarrow\ \  0=\Delta_+-\Delta_- = 108 v \Lambda_0^3\ .
\ee
This requires $v=0$ and
\be
u^3=\frac{27\Lambda_0^6}{4}\ .
\label{symlimit}
\ee
In particular, for the root $u=3\Lambda_0^2/2^{\frac23}$, the curve has the form
\be
y^2=\left( x -2^{\frac23} \Lambda_0 \right)\left( x +2^{\frac23} \Lambda_0 \right)
\left( x -2^{-\frac13} \Lambda_0 \right)^2\left( x +2^{-\frac13} \Lambda_0 \right)^2
\ee
On this singularity, there are two dyonic states becoming massless, and ${\rm Im}\big( a_{D1}\big)={\rm Im}\big(a_{D2}\big)=0$.
This is the dominant saddle-point as it has the least action.
The free energy $F=-\ln Z$ for this theory is therefore given by
\be
F=- \frac{18}{2^{\frac23}}  R^2 \Lambda_0^2\ .
\label{freelow}
\ee
The saddle-point at $u=3\Lambda_0^2/2^{\frac23}$, $v=0$, will describe the low-energy limit of the $SU(3)$ theories with $N_f<2N$ massive hypermultiplets considered below.


\subsection{$SU(3)$ gauge group with $N_f=4$}

A detailed discussion of singularities and superconformal points in this case was given in \cite{Eguchi:1996vu}.
The hyperelliptic curve is
\begin{equation}
\label{nfcuatro}
    y^2=Q_+ Q_- ,\qquad Q_\pm = x^3-ux-v+\frac14 \Lambda^2(x+4M)  \pm \Lambda (x+M)^2\ .
\end{equation}
The branch points are the zeros of $Q_\pm$, which we shall denote by $\{ x^+_k, x^-_k\} ,\ k=1,2,3 $.
In the classical limit, they behave as $x^\pm_k\sim c_k\ v^{\frac13}+O(v^0)$. They are labelled such that
$x^+_k-x^-_k =\frac{2\Lambda}{3}+O(v^{-\frac13})$.
The period integrals $a_k$ are then defined in terms of the contours encircling $x^+_{k+1}$,  $x^-_{k+1}$, whereas the  $a_{Dk}$ integrals
loop around $x^+_{1}$ and $x^{+}_{k+1}$, with $k=1,2$. \footnote{We adopt the same choice of one-cycles as in \cite{Hanany:1995na}. A general discussion of the monodromies in the classical limit for any $N$, $N_f$ can be found in this paper.} 

The discriminant is given by
\begin{equation}
    \Delta =c\Delta_s^4 \Delta_{m+}\Delta_{m-}\ ,\qquad \Delta_s=  M^3-Mu+v-\frac34 M\Lambda^2\ ,\
\end{equation}
\begin{equation}
    \Delta_{m\pm} = 4u_\pm^3 -27(v_\pm \mp \frac{\Lambda}{3}u_\pm)^2\ ,
\end{equation}
with\footnote{Here we correct a typo in (36), (37) of \cite{Eguchi:1996vu}.} 
\begin{equation}
    u_\pm = u\mp 2\Lambda M+\frac{\Lambda^2}{12}\ ,\qquad v_\pm=v\mp \Lambda M^2-\Lambda^2M\pm 
    \frac{\Lambda^3}{27}\ .
\end{equation}
There is a critical point at $\Lambda_c=3 M$, which we will describe below.

In order to identify the saddle-point which is relevant to the strong coupling phase $\Lambda>3M$ 
we first solve $\Delta_s=0$. This gives
\begin{equation}
    v=v_s= Mu+\frac34 M\Lambda^2 - M^3\ .
\label{vvssfour}
\end{equation}
At $v=v_s$, two branch points join giving
\begin{equation}
    y^2\bigg|_{v=v_s}=(x+M)^2 (x-x_1) (x-x_2)(x-x_3) (x-x_4)\ ,
\end{equation}
with
\bea
x_{1,2} &=& \frac12 \left(M-\Lambda \pm \sqrt{4u-3M^2-6M\Lambda }\right)\ ,\   
\nonumber\\
x_{3,4} &=& \frac12 
\left(M+\Lambda \pm \sqrt{4u-3M^2+6M\Lambda }\right) . 
\label{raicesa}
\eea
Note that $x_{1,2}$ and $x_{3,4}$ are exchanged under $\Lambda\to -\Lambda $.
In order to solve the two saddle-point conditions,
\be
{\rm Im}\big(a_{D1}\big)=0\ ,\quad  {\rm Im}\big(a_{D2}\big)=0\ ,
\ee
 we need another cycle
shrinking to zero.
From the form of the roots \eqref{raicesa}, we see that at
\be
u_s=\frac34 M^2 +\frac 32 M\Lambda\ ,
\label{uussfour}
\ee
the branch points $x_1$ and $x_2$ collide, giving the behavior
\be
  y^2\bigg|_{u_s, v_s}=(x+M)^2 \left( x-\frac12 (M-\Lambda) \right)^2 (x-x_3) (x-x_4)\ ,
\ee
with
\be
x_{3,4}= \frac{1}{2} \left( M+\Lambda\pm 2\sqrt{3M\Lambda }\right) \ .
\ee
There are two other possible singular points: 
\smallskip

\noindent 1) $u_1=\frac34 M^2 -\frac 32 M\Lambda$. This solves $\Delta_s=\Delta_{m-}=0$.
As it will be clear from the discussion below, this does not match
the saddle-point of the weak coupling phase at the critical point.

\noindent 2) $u_2=3 M^2 +\frac 14 \Lambda^2$. This gives a behavior $y^2\approx (x+M)^4$. 
In the classification of \cite{Eguchi:1996vu}, it is a class 3 theory involving superconformal sectors. 
But neither does this singularity  match the saddle-point of the
weak coupling phase at the critical point.

\smallskip

 We must now check that the saddle-point equations are satisfied at $(u_s,v_s)$. Which cycles contract
at $u=u_s$ and $v=v_s$ depend on whether we are in the  subcritical or supercritical regime. Explicitly,
we find
\bea
&& \Lambda> 3M:\ \qquad x^+_2 =x^+_3=\frac12 (M-\Lambda)\ ,\qquad x_1^+=x^-_2=-M\ ,
\nonumber\\
&& \Lambda< 3M:\ \qquad x^+_1 =x^+_3=\frac12 (M-\Lambda)\ ,\qquad x_2^+=x^-_2=-M\ .
\eea
Hence 
\bea
&& \Lambda> 3M:\ \qquad  \ a_{D1}=a_{D2}\ ,\ \ \ a_{D1}=a_1\  ,
\nonumber\\
&& \Lambda< 3M:\ \qquad \ a_{D2}=0\ ,\ \ \ a_1=0\ .
\eea
Consider first $\Lambda>3M$. The condition $a_{D1}=a_1$ (which holds modulo an additive real  mass term),
solves ${\rm Im} (a_{D1})=0$ because $a_1$ is real. Since $a_{D2}=a_{D1}$, then ${\rm Im} (a_{D2})=0$ as well,
and the saddle-point conditions are satisfied. We have checked this by integrating the
one-form $\lambda $ \eqref{merom} from $x_1^+$ to $x_2^+$ -- to compute $a_{D1}$ --
and from  $x_1^+$ to $x_3^+$ -- to compute $a_{D2}$.
On the other hand, when $\Lambda<3M$,  ${\rm Im}\big(a_{D1})$ is non-vanishing.
Therefore the singularity at $u=u_s$ and $v=v_s$ does not solve the saddle-point equations when  $\Lambda<3M$.
Note that the required matching with pure $N_f=0$ SYM  at $M\to \infty $ already implies
that in the weak coupling phase
the dominant saddle-point must correspond to a different singularity, since $(u_s,v_s)$ moves to infinity in the $M\to\infty $ limit.
The singularity in the moduli space that dominates the partition function integral must be different.
This jumping from one singular point to another one in crossing $\Lambda=3M$ in turn implies a phase transition.

Sitting first  on $u=u_s$, $v=v_s$, as the coupling $\Lambda/M$ is decreased from strong values, we get to a critical point $\Lambda =3M$
where five branch points coincide, so that the behavior is
\be
  y^2\bigg|_{u_s,v_s,\Lambda_c}=(x+M)^5 (x-5M)  \ .
\ee 
The curve undergoes maximal degeneration.
This corresponds to a class 4 theory representing a strongly interacting superconformal field theory
 (it includes operators of scaling dimensions $\frac23,\ \frac43,\ 2$, see \cite{Eguchi:1996vu} and section 4.5 for more details).
Let us now consider the weak coupling phase $\Lambda< 3M$. 
The other possible singularities occur when $\Delta_{m+}=\Delta_{m-}=0$. Considering the equation $\Delta_{m+}-\Delta_{m-} =0$,
we find a linear equation for $v$. It gives the solution
\be
v=\tilde v \equiv \frac{4 \left(25 \Lambda ^2 M^3+12 M u^2+5 \Lambda ^2 M u\right)}{36 u-\Lambda ^2+108
   M^2}\ .
\ee
Substituting $v=\tilde v$ into $\Delta_{m+}=0$, we obtain a cubic polynomial equation for $u$:
\be
u^3-\frac{\Lambda^2}{18} u^2+\left( \frac{\Lambda^4}{6^4} -\frac32 \Lambda^2M^2\right) u+\frac{M^2\Lambda^4}{216}-\frac{27}{4} M^4\Lambda^2=0\ .
\ee
By construction, 
the zeros automatically match the solutions \eqref{symlimit} occurring in the $N_f=0$ theory  when $M\to\infty$ at fixed $\Lambda_0^3= \Lambda M^2$ (it is automatic because the curve  \eqref{nfcuatro} of the $N_f=4$ theory reduces to  the $N_f=0$ curve \eqref{nfcero} in this limit).
The  question is which of the three roots of the cubic equation is the one that matches
$(u_s,v_s)$, given in \eqref{vvssfour}, \eqref{uussfour},  at the critical point $\Lambda=3M$. There is one real root $\tilde u$ and two complex conjugate roots
in the whole interval $0<\Lambda<3M$.
A numerical inspection shows that it is the real root  the one that matches $u_s,v_s$  at $\Lambda=3M$. The complex conjugate roots give rise to complex values for $u,v$ at $\Lambda=3M$. It is also the real root the one that approaches $(u,v)=(\frac{3}{2^{\frac23}} \Lambda^{\frac23} M^{\frac43},0) $ at $M \to \infty$.
We thus find
that the real root $(\tilde u, \tilde v)$ is the only one that matches $(u_s,v_s)$  at $\Lambda=\Lambda_c$ and also the one that is consistent with the renormalization group flow to pure SYM. The contracted cycles are as follows:
  \be 
{\rm At}\ (u,v)=(\tilde u,\tilde v):\qquad  x^{-}_{2}=x^{-}_{3}\ ,\quad x^{+}_1=x^+_{3}\ .
\ee
$x^{+}_1=x^+_{3}$ implies $a_{D2}=0 $ and $ x^{-}_{2}=x^{-}_{3}$, together with
$a_{D2}=0$, implies $a_{D1}=a_1+a_2$. The roots are all real and $a_1$, $a_2$ are real.
Thus the saddle-point equations ${\rm Im}\big(a_{D1}\big)={\rm Im}\big(a_{D2}\big)=0$
are satisfied (as it is easy to check by explicit integration using \eqref{merom}). On the other hand, when $\Lambda>3M$, at $(\tilde u,\tilde v)$ one
still has $x^{-}_{2}=x^{-}_{3}$. However, now $x^{+}_2=x^+_{3}$.
This implies $a_1=a_2$, $a_{D1}=a_{D2}$.
As a result,  $(\tilde u,\tilde v)$ does {\it not} solve the saddle-point equations in the strong coupling phase: the only consistent 
solution in the strong coupling phase is $(u_s,v_s)$, as described above. Thus the saddle-point jumps from the singularity
at $(\tilde u,\tilde v)$  to the singularity
at $(u_s,v_s)$ in crossing  the critical point.

In conclusion, we have a complete characterization of the two phases of the theory.
The theory has two quantum phases and it undergoes a phase transition at $\Lambda=3M$.
The free energy in each phase is a function of $\Lambda/M$ and it is obtained from the prepotential
as $F=-R^2 {\rm Re} (4\pi i {\cal F})$ computed at $(u_s,v_s)$ in the strong coupling phase and at $(\tilde u,\tilde v)$ in the weak coupling phase.
Substituting these values for $u,v$ into the formula for the periods, one may compute $a_1,\ a_2$ as  functions of $\Lambda/M$ and
hence the free energy $F(\Lambda/M)$. 
In the massless limit -- representing the strong coupling limit  $\Lambda/M\to\infty$ -- we may again make use of the formula (\ref{freeenergy}). This gives
\be
F\bigg|_{M\to 0}=-2R^2\Lambda^2 {\rm Re}\big( u_s\big) \to 0\ .
\ee
In the weak coupling  limit $\Lambda\ll M$, the free energy flows to \eqref{freelow}.
For the general dependence $F(\Lambda/M)$, in the present theory with $SU(3)$ gauge group, formulas are much more complicated than those seen in the rank 1, $SU(2)$ case
(see  \cite{Ohta:1996fr,Masuda:1996xj} for discussions).
Nonetheless, it is possible to understand some features of the critical behavior, which
we shall discuss in section 4.5.

\subsection{$SU(3)$ gauge group with $N_f=2$}

The hyperelliptic curve describing SQCD with two massive flavors of equal masses is
\begin{equation}
    y^2=Q_+(x) Q_-(x)\ ,\qquad
\end{equation}
where now 
\be
 Q_\pm(x)\equiv (x^3-u x-v)\pm  \Lambda^2 (x+M)\ .
\ee
The zeros of $Q_\pm $ are the branch points appearing at $\{ x^{+}_{k}, x^{-}_{k}\}$, $k=1,2,3$. We label them in such a way that in the classical  limit, where $v$ is large, $x^{+}_{k}-x^{-}_{k}=O(v^{-1/3}) $.
The period integrals $a_k$ are then defined in terms of the contours encircling $x^+_{k+1}$,  $x^-_{k+1}$, whereas the  $a_{Dk}$ integrals
loop around $x^+_{1}$ and $x^{+}_{k+1}$, with $k=1,2$. 

Singularities arise at zeros of the discriminant, which now reads
\begin{equation}
    \Delta =\Delta_s^2 \Delta_{m+} \Delta_{m-}\ ,\qquad \Delta_s=  M^3-Mu+v\ .
\end{equation}
Here  $\Delta_{m\pm} $ are the discriminants of $Q_\pm$, representing the monopole singularities,
\be
\Delta_{m\pm} = 4\left( u\mp \Lambda^2 \right)^3 - 27\left( v\mp M\Lambda^2 \right)^2
\ .
\ee
The theory has two phases, with a critical point at $\Lambda_c = \frac{3}{2\sqrt{2}} M $.
The strong coupling phase $\Lambda>\Lambda_c$ is found by 
first solving $\Delta_s=0$. This gives
\be
v=v_s= Mu-M^3\ .
\ee
At this $v$, two branch points coincide and the curve takes the form
\be
  y^2\bigg|_{v=v_s}=(x+M)^2 (x-x_1)(x-x_2)(x-x_3)(x-x_4)
  \  ,
\ee
\be
x_{1,2}=\frac12 \left(M \pm \sqrt{4u-3M^2-4\Lambda^2 }\right)\ ,\quad x_{3,4}=\frac12 \left(M \pm \sqrt{4u-3M^2+4\Lambda^2 }\right)\ .\
\ee
As we have two saddle-point equations, ${\rm Im}\big(a_{D1}\big)=0 ,\  {\rm Im}\big(a_{D2}\big)=0$,
in addition, we must demand that $\Delta_{m+} \Delta_{m-}$ vanishes. This will lead to a second  cycle contracting to zero.
$\Delta_{m+}=0$ has solutions $u_1=3M^2+\Lambda^2$ and 
\be
u_s=\frac34 M^2+ \Lambda^2\ .
\label{uusin}
\ee
The solutions of $\Delta_{m-}=0$ are similar with the change $\Lambda^2\to -\Lambda^2$. 
As it is clear from the analysis below, by demanding continuity with the weak coupling phase, one can rule out the other possible singularities arising from $\Delta_{m-}=0$. 

The singularity at $u_1$ gives a behavior $y^2\sim (x+M)^3$. In the classification of \cite{Eguchi:1996vu}, this corresponds to a
class 4 superconformal field theory. It does not solve the saddle-point equations.

On the other hand, at $u_s,v_s$, we find the behavior
\be
  y^2\bigg|_{u_s,v_s}=(x+M)^2 \left(x-\frac{M}{2} \right)^2 
 (x-x_3)(x-x_4) \ ,\qquad x_{3,4}=\frac{M}{2}\pm \sqrt{2}\Lambda\ .
\ee
We must now inspect what  cycles shrink.
The answer depends on whether $\Lambda>\Lambda_c$ or  $\Lambda<\Lambda_c$,
as follows:
\bea
&& \Lambda>\frac{3}{2\sqrt{2}} M:\qquad x^{+}_2=x^-_{3}=-M\ ,\quad x^+_{1}=x^+_{3}=\frac{M}{2}\ , 
\nonumber\\
&& \Lambda<\frac{3}{2\sqrt{2}} M: \qquad x^+_{2}=x^-_{2}=-M\ ,\quad x^+_{1}=x^+_{3}=\frac{M}{2}\ .
\eea
Thus, for $\Lambda>\Lambda_c $ we have $a_{D2}=0$ and $-a_{2}+a_{D1}= a_{D2}$ (modulo a real mass residue), i.e.
$a_{D1}=a_2$.
This solves the saddle-point equations, as one can also explicitly check by numerical integration of \eqref{merom} that ${\rm Im}\big( a_{D1}\big)=0$.
However, for $\Lambda<\Lambda_c $, we get  $a_{D2}=0$ and $a_1=0$, which
does not solve the remaining saddle-point equation ${\rm Im}\big( a_{D1}\big)=0$.
Therefore the partition function must be controlled by a different singularity, implying a phase transition.

When $\Lambda $ reaches the critical point at  $\Lambda_c = \frac{3}{2\sqrt{2}} M $, the degeneracy
increases and
\be
  y^2\bigg|_{u_1,v_s,\Lambda_c}=(x+M)^3 \left(x-\frac{M}{2} \right)^2 
 (x-2M) \  .
\ee
The critical point does not have maximal criticality (which would be $y^2\sim (x+M)^4$ for $N=3,\ N_f=2$) but it still
describes a (class 4) strongly interacting superconformal field theory \cite{Eguchi:1996vu}. 

The partition function in the weak coupling phase must match the partition function of pure SYM, upon making the identification $\Lambda_0^3=\Lambda^2 M$ and taking the limit $M\to \infty $, $\Lambda\to 0$.
Comparing with the results of section 4.1, one sees that the relevant saddle-point is found  by choosing a  solution of the monopole singularity $\Delta_{m+}=\Delta_{m-} =0$ for $u,v$. 
It must be noted that the singularity \eqref{uusin} of the strong coupling phase, arising from $\Delta_s=0$, moves to infinity as $M\to\infty$, which is another reason why it cannot describe the saddle-point in the weak coupling phase.

To solve $\Delta_{m+}=\Delta_{m-} =0$, we first consider the linear equation for $v$ obtained from $\Delta_{m+}-\Delta_{m-} =0$.
This gives the solution:
\be
v=\tilde v\equiv \frac{2 \left(\Lambda ^4+3 u^2\right)}{27 M}\ . 
\ee
Substituting  $\tilde v$ into $\Delta_{m+}=0$ we get a quartic equation for
$u$. There are two real roots and two complex conjugate roots.
In order to identify which root represents the saddle-point, we must now use the matching conditions, namely
consistency with the  renormalization group flow to the $N_f=0$ theory and continuity at the critical point.
Under the renormalization group flow $M\to \infty $, $\Lambda\to 0$ with fixed $\Lambda_0^3=\Lambda^2 M$,
the relevant singularity $(\tilde u,\tilde v)$  must approach the saddle-point found in section 4.1, $(\frac{3}{2^{\frac23}} \Lambda_0^2,0)$. 
One real root moves to infinity in the $M\to\infty$ limit, so it is ruled out.
The remaining real root $\tilde u$ approaches the $N_f=0$ saddle point and it is the only one that matches $u_s$  at $\Lambda=\Lambda_c$, where we also have $\tilde v=v_s$. On this singularity, we find
 \be 
(u,v)=(\tilde u,\tilde v):\qquad  x^{-}_{2}=x^{-}_{3}\ ,\quad x^{+}_1=x^+_{3}\ , 
\ee
$x^{+}_1=x^+_{3}$ implies $a_{D2}=0 $ and $ x^{-}_{2}=x^{-}_{3}$, together with
$a_{D2}=0$, implies $a_{D1}=a_1+a_2$. The periods  $a_1$, $a_2$ are real and one has
 ${\rm Im}\big(a_{D1}\big)={\rm Im}\big(a_{D2}\big)=0$, as can  be checked by numerically computing the period integrals.

The singularity at $(\tilde u,\tilde v)$ also represents a saddle-point in the
strong coupling phase $\Lambda>\Lambda_c$, because it still holds that
$x^{-}_{2}=x^{-}_{3} ,\  x^{+}_1=x^+_{3}$.
Thus, at $\Lambda>\Lambda_c$, we have two competing saddle-points: $(\tilde u,\tilde v)$
and $(u_s,v_s)$. To see which one is dominant, we look at the limit $\Lambda\gg \Lambda_c$. This is the massless limit,
where the free energy is computed by (\ref{freeenergy}). In this limit, $u_s\to \Lambda^2$
and $\tilde u\to \frac{i\Lambda^2}{\sqrt{3}}$. Since $F\to -4R^2 {\rm Re}(u)$, it follows that the action is less at $(u_s,v_s)$ , the contribution from 
$(\tilde u,\tilde v)$ being exponentially suppressed.
Therefore $(u_s,v_s)$ is the relevant saddle-point in the strong coupling phase.
In the strong coupling limit the free energy is  given by the remarkably simple formula:
\be
F \bigg|_{\Lambda\gg M}= -4 R^2\Lambda^2\ .
\ee

Summarizing,  the theory undergoes a phase transition as
$\Lambda/M$ is increased from 0 to infinity, with a critical point at $\Lambda_c = \frac{3}{2\sqrt{2}} M $. 
The critical point is a non-maximally singular point described by a class 4 strongly interacting superconformal field theory.

\subsection{$SU(3)$ gauge group with $N_f=3$}

For three flavor hypermultiplets of mass $M$, the hyperelliptic curve takes the form:
\begin{equation}
    y^2=\left( x^3-ux-v+\frac14 \Lambda^3 \right)^2 -\Lambda^3 (x+M)^3\ .
\end{equation}
In this case of odd $N_f$, there is no factorization and branch points are now given by the zeros  of a sixth-order polynomial.
Nonetheless, it is possible to completely determine the phases and the free energy.
The discriminant is now given by
\begin{equation}
    \Delta = c\Delta_s^3 \Delta_m\ ,\qquad \Delta_s=  M^3-Mu+v-\frac14 \Lambda^3\ .
\end{equation}

In order to identify the correct cycles for the periods $a_1, a_2$ and $a_{D1}$, 
$a_{D2}$, we first consider the classical limit of $v $ large, where the curve takes the form
\be
y^2 \approx \left( x^3-v \right)^2 -\Lambda^3 x^3\ ,\qquad v\gg \Lambda^3,\ M^3\ .
\ee
The branch points are located at 
\be
x^\pm_k =\left(\frac{1}{2} \left(2v+\Lambda ^3 \pm \sqrt{\Lambda ^6+4 \Lambda ^3 v}\right)\right)^{\frac13}\ ,\ \qquad
\ee
where $k=1,2,3$ corresponds to the three cubic roots.
Similarly as in previous cases,  the cycles defining the period integrals $a_k$ 
loop around $x^+_{k+1}$,  $x^-_{k+1}$, whereas  $a_{Dk}$ integrals
are defined in terms of cycles around $x^+_{1}$ and $x^{+}_{k+1}$, with $k=1,2$. 
It will be shown below that this theory has two phases with a critical point at 
\be
\Lambda_c = 2^{\frac23} M\ .
\ee

In the strong coupling phase $\Lambda>\Lambda_c$, as in previous cases the relevant saddle-point is found by first solving
$\Delta_s=0$. This gives
\be
v=v_s=\frac{\Lambda^3}{4}+ M u -M^3\ .
\ee
For $v=v_s$, one finds the behavior
\begin{equation}
    y^2\bigg|_{v=v_s}=(x+M)^2 p_4(x)\ ,\quad p_4(x)=\left(M^2-M x-u+x^2\right)^2-\Lambda ^3 (M+x).
\end{equation}
To find another double degeneracy, we demand that the discriminant of $p_4(x)$ vanishes.
This gives the cubic polynomial equation
\be
\label{qubi}
u^3-\frac{9 M^2 u^2}{2}+\frac{27}{16} u \left(3 M^4-\Lambda ^3 M\right)-\frac{27}{256} \left(-\Lambda ^6+16 M^6-44 \Lambda ^3 M^3\right)=0 \ .
\ee
As in the previous examples, the choice of the relevant root
is dictated by the requirement of continuity at the critical point. This will select only one root, that we call $u_s$, as the relevant saddle-point.
In the massless limit, the roots are 
\be
\label{noconozco}
M=0:\qquad u_1^{(0)}= -\frac{3}{2^{\frac83}}\Lambda^2\ ,\quad u_2^{(0)}= \frac{3}{2^{\frac83}}\  e^{-\frac{i\pi}{3}}\Lambda^2\ ,\quad u_3^{(0)}= \frac{3}{2^{\frac83}}\ e^{\frac{i\pi}{3}}\Lambda^2 .
\ee
$u_s$ is defined as the zero of \eqref{qubi} that approaches $u_3^{(0)}$ as $M\to 0$.

On the singularity at $(u_s,v_s)$, the curve has the form
\begin{equation}
    y^2\bigg|_{v_s,u_s}=(x+M)^2 \left( x-x_0\right)^2 (x- e_1)(x- e_2)\ .
\end{equation}

The cycles that contract to zero size can be found by numerical inspection
of the behavior of the six branch points.
We find that, on the singularity $(u_s,v_s)$, there are two massless dyons and the equations ${\rm Im}(a_{D1})=  {\rm Im}(a_{D2})=0$ are satisfied, provided  $\Lambda> 2^{\frac23} M$.
Thus the singularity at $(u_s,v_s)$ represents a saddle-point in the supercritical regime.
In the strong coupling limit, the free energy can be computed from \eqref{freeenergy}. We get
\be
F\bigg|_{\Lambda\gg M}=-3R^2\Lambda^2 {\rm Re}\big( u_s\big) =-\frac{9 R^2\Lambda^2 }{2^{\frac{11}{3}}}\ .
\ee

At the critical point $\Lambda_c = 2^{\frac23} M$, the curve becomes more degenerate
\begin{equation}
\label{cricu}
    y^2\bigg|_{v_s,u_s,\Lambda_c}=x^2 (x+M)^3 (x-3M)\ .
\end{equation}
In the classification of \cite{Eguchi:1996vu}, this is a class 2 theory, representing an interacting superconformal field theory.
It was noted in \cite{Eguchi:1996vu} that these fixed points describe an universal class of SCFT  appearing 
for any $SU(N)$ with odd $N_f<2N$; in particular, the dimensions
of the relevant operators are independent of $N$.


The singularity at $v=v_s$, arising as a solution of $\Delta_s=0$, moves to infinity in the $M\to \infty$ limit. 
This already implies that the  saddle-point of the subcritical phase  $\Lambda < 2^{\frac23} M$
must be different, since $v=v_s$ cannot satisfy the matching condition with pure SYM.
The relevant saddle-point describing the weak coupling phase $\Lambda< 2^{\frac23} M$ is obtained by
solving $\Delta_m=0$ only. To have a double degeneracy, we must
also demand that the discriminant of $\Delta_m$ (viewed as a polynomial in $v$) vanishes.
This gives  the solution
\be
\tilde u^3=\frac{27 M^3 \Lambda^3}{4}\ .
\ee
This matches the  solution (\ref{symlimit}) in the $M\to \infty $ limit at fixed $\Lambda_0^2 =M\Lambda $.
In fact, we must take the real cubic root.
Substituting $\tilde u$ into $\Delta_m=0$, we find a quartic equation for $v$ with
a double root at
\be
\tilde v =\frac{3M\Lambda^2}{2^{\frac43}}\ .
\ee
Note that $\tilde v\to 0$ in the $M\to \infty $ at fixed $\Lambda_0^2=\Lambda M$, in consistency
with  the renormalization group flow to pure Super Yang-Mills theory.
The singularity at $(\tilde u,\tilde v)$  gives rise to the correct structure for $y^2$,
\be
y^2 =(x-x_1)^2(x-x_2)^2 (x-x_3)(x-x_4)\ ,
\ee
\be
x_{1,2}= -\frac{\Lambda \pm \sqrt{\Lambda  \left(2^{8/3} M-3 \Lambda
   \right)}}{2^{5/3}}
\ ,\quad 
x_{3,4}=\frac{\Lambda }{2^{2/3}}\pm 2^{2/3} \sqrt{\Lambda M} \ .
\ee
Note that the branch points are real in this phase, since $\Lambda < 2^{\frac23} M$.

As $\Lambda $ is increased from 0, we get to the superconformal point $\Lambda = 2^{\frac23} M$ where the curve takes the form
\be
y^2 =x^2 (x+M)^3 (x-3M)\ ,
\ee
in agreement with (\ref{cricu}). Furthermore, one can check that no continuous matching is possible between
the other roots $u_{1,2}$ of (\ref{qubi}) and any solution that matches pure super Yang-Mills theory at low energies.

In conclusion, $SU(3)$ SQCD with three fundamental hypermultiplets of equal mass undergoes a phase transition at $\Lambda_c =2^{2/3}M$. The partition function in the strong coupling $\Lambda>\Lambda_c$ phase  is dominated by  the singularity $(u_s,v_s)$, whereas in the $\Lambda<\Lambda_c$ phase is dominated by the singularity $(\tilde u,\tilde v)$.

Finally, we note the existence of a special point in the strong coupling phase, 
which occurs at $\Lambda =2M$, where the branch points simplify, with double degeneracy at $x=-M$ and $x=M$. The curve would get a cusp singularity if one sits
on the singularities $u_1$ or $u_2$, corresponding to the roots $u_{1,2}^{(0)}$ in the massless limit \eqref{noconozco}. Although these singularities are ruled out by the requirement of 
continuity of the partition function at $\Lambda_c =2^{2/3}M$,
it would be interesting to understand if the presence of this special point is reflected into some non-analytic behavior of the free energy.


\subsection{Critical behavior}

The critical behavior of the free energy can be understood
from the analysis of perturbations about the fixed point. 
We will follow the analysis of   \cite{Eguchi:1996vu}, to which we refer for further details.
We start with the case $N_f=3$. 
The critical point of the phase transition is described by a class 2 superconformal theory, arising for odd $N_f$, where
the singularity is of the form $y^2\approx (x+M)^{N_f}$. In this case the scaling
dimensions of the perturbations $t_j$ are independent of $N$, depending only on $N_f$,
\be
[t_j]=\frac{N_f}{2}-j\ ,\qquad j=0,1,...,\frac12(N_f-1)\ .
\ee
Consider a period $a_k$ or $a_{Dk}$ that vanishes at the critical point.
Periods have  scaling dimension equal to 1. 
This imply that they have the behavior
\be
a_k,\ a_{Dk} \approx t_j^{\alpha_j}\ ,\qquad \alpha_j=\frac{1}{[t_j]}\ .
\ee
The leading critical exponent originates from the perturbation associated with the chiral operator of highest scaling dimension
$[t_0]=N_f/2$.
For this perturbation, we thus find
\be
N_f=3:\qquad a_k\approx t_0^{\frac{2}{3}},\quad a_{Dk}\approx t_0^{\frac{2}{3}}\ .
\ee
This is the same behavior as  in the $SU(2)$ case with $N_f=3$, as expected since
class 2 theories represent a universality class of SCFT with global $SU(N_f)$ flavor symmetry, which are independent of $N$.
The perturbation $t_0$ can be identified with $v_s-v_c$ and it vanishes at the critical point as $t_0\sim \Lambda-\Lambda_c$. Since the leading non-analytic behavior of the prepotential is ${\cal F}\sim t_0^{\frac43}$,  the susceptibility  $\chi =-\partial^2_{\Lambda} F\to\infty$ as $\Lambda\to\Lambda_c$.
Therefore the theory undergoes a second-order phase transition, like in the $SU(2)$ case.

Let us now consider the cases $N_f=2,\ 4$. We found that the critical points correspond to 
class 4 theories, which arise for even $N_f$.
These are defined by having a singularity of the form
\be
y^2\approx (x+M)^{p+N_f}\ ,\qquad 0<p\leq N-N_f/2\ .
\ee
In the case $N_f=4$, we found maximal criticality, i.e. $p=N-N_f/2=1$, 
$y^2\approx (x+M)^5$.  
The underlying low-energy theory  was conjectured \cite{Gaiotto:2010jf} to be described by two superconformal field theories
coupled by an infrared free, magnetic $SU(2)$ gauge theory (see also \cite{Giacomelli:2012ea}).
The $\beta $ function of the  $SU(2)$ gauge theory may induce logarithmic terms in the prepotential, in which case computing
critical exponents is delicate.
We can however make an estimate of the leading critical behavior of the periods. 
Following  \cite{Eguchi:1996vu}, we consider the perturbation
\be
y^2\approx \big((x+M)^3-t_0\big) \left( x+M\right)^{N_f/2}\ .
\ee
The one-form $\lambda $ (\ref{merom}) behaves as
\be
\lambda\approx t_0^\alpha\ ,\qquad \alpha\equiv \frac{5-N_f/2}{6}=\frac12 \ .
\ee
Thus the vanishing periods  have the behavior
\be
 N_f=4:\qquad a_k,\ a_{Dk} \approx t_0^{\frac12}\ ,\qquad t_0\sim \Lambda-\Lambda_c\ .
\nonumber
\ee

Finally, let us  consider $N_f=2$. In this case we found $y^2\approx (x+M)^3$, corresponding to
lower ($p=1$) criticality (for two flavors, maximal criticality would be achieved by $p=2$). 
The low-energy theory for this non-maximally singular point has not yet been studied. By the same arguments of \cite{Gaiotto:2010jf}, it is plausible that  the low energy theory  involves non-conformal sectors with running coupling, which may again lead to logarithmic terms in the prepotential around this point.  As in the $N_f=4$ case, we will here provide an estimate
of the critical behavior of the periods. We found $a_{D2}=a_1=0$, $a_2=a_{D1}$ at the critical point.
The scaling dimensions of the perturbations around this singularity are given by
  \cite{Eguchi:1996vu}
\be
[t_j]= \frac{2(N-j)}{p+2}= \frac{2(3-j)}{3}\ ,\qquad j=0,...,N-1\ .
\ee
Thus the highest  dimension is $[t_0]=2$, which implies   the leading
near-critical behavior
\be
 N_f=2:\qquad a_1\approx t_0^{\frac12}\ ,\ \ \ a_{D2}\approx t_0^{\frac12}\ ,\ \ \ a_{D1}-a_2\approx t_0^{\frac12}\ .
\nonumber
\ee
Thus, around the critical point, the periods scale in the same way as in the theory with four flavors. 
It would be interesting to clarify the critical behavior of the prepotential.

\section{Summary}

This concludes our survey over the phase structure of ${\cal N}=2$ supersymmetric QCD in four dimensions with gauge groups
$SU(2)$ and $SU(3)$. Our method was based on an exact saddle-point evaluation of the partition function and
the observation that the saddle-point corresponds to a singularity where $N-1$ dyons become massless. This fixes $N-1$ moduli parameters $s_k$ leaving a discrete set of solutions $\{ s_k\} $.
The dominant saddle-point in each phase was identified by matching conditions and  by the explicit computation
of the action in the two limits, $\Lambda\gg M$ and $\Lambda\ll M$. In the case $\Lambda\ll M$, the theory flows
to pure super Yang-Mills theory and one of the matching conditions requires the saddle-point calculation
to reproduce the pure SYM  partition function in this limit.
The second matching condition requires continuity at the critical point, which implies that
the moduli of the weak and strong coupling phases must coincide at this point. 
This, along with the  least-action principle in the  $\Lambda\gg M$ limit, uniquely determine the saddle points
of the two phases in the examples considered here.

The critical point of the phase transition corresponds to a singular point
in the Coulomb branch and is described by interacting superconformal theories, whose
 origin in terms of the emergence of mutually non-local massless states is well understood \cite{Argyres:1995jj,Argyres:1995xn,Eguchi:1996vu,Gaiotto:2010jf}.
The scaling dimensions of chiral operators at the fixed point
dictate the behavior of the free energy at criticality and thus the order of the phase transition.
We have argued that the phase transition is second order in the $SU(2)$ case for $N_f=1,2,3$ and in the $SU(3)$ case with $N_f=3$.

One interesting problem is understanding the phase structure in the complex $\Lambda$ plane and the
convergence properties of the expansion in powers of $\Lambda/M$.
Another interesting problem concerns the case of the
${\cal N}=2^*$ $SU(N)$ theory, whose partition function in the weak coupling phase was calculated exactly in \cite{Hollowood:2015oma}. A striking feature of this theory is the resurgence of the Wigner semicircle distribution  for the eigenvalue density in an coarse-grained form \cite{Russo:2013kea,Buchel:2013id}, occurring in the strong coupling limit, where results can be compared with holography. 
Perhaps this feature can also be elucidated by appropriate matching conditions, despite the more complicated phase structure.

\section*{Acknowledgments}

We acknowledge financial support from projects 2017-SGR-929, MINECO
grant FPA2016-76005-C.


\begin{thebibliography}{99}

\bibitem{Seiberg:1994rs}
  N.~Seiberg and E.~Witten,
  ``Electric - magnetic duality, monopole condensation, and confinement in N=2 supersymmetric Yang-Mills theory,''
  Nucl.\ Phys.\ B {\bf 426} (1994) 19
   Erratum: [Nucl.\ Phys.\ B {\bf 430} (1994) 485]
  [hep-th/9407087].


\bibitem{Seiberg:1994aj} 
  N.~Seiberg and E.~Witten,
  ``Monopoles, duality and chiral symmetry breaking in N=2 supersymmetric QCD,''
  Nucl.\ Phys.\ B {\bf 431}, 484 (1994)
  [hep-th/9408099].


\bibitem{Nekrasov:2002qd}
  N.~A.~Nekrasov,
  ``Seiberg-Witten Prepotential From Instanton Counting,''
  Adv.\ Theor.\ Math.\ Phys.\  {\bf 7}, 831 (2004)
  [arXiv:hep-th/0206161].



\bibitem{Nekrasov:2003rj}
  N.~Nekrasov and A.~Okounkov,
  ``Seiberg-Witten theory and random partitions,''
  arXiv:hep-th/0306238.

\bibitem{Pestun:2007rz} 
  V.~Pestun,
  ``Localization of gauge theory on a four-sphere and supersymmetric Wilson loops,''
  Commun.\ Math.\ Phys.\  {\bf 313}, 71 (2012)
  [arXiv:0712.2824 [hep-th]].


\bibitem{Russo:2013kea} 
  J.~G.~Russo and K.~Zarembo,
  ``Massive N=2 Gauge Theories at Large N,''
  JHEP {\bf 1311}, 130 (2013)
  [arXiv:1309.1004 [hep-th]].

\bibitem{Russo:2013sba} 
  J.~G.~Russo and K.~Zarembo,
  ``Localization at Large N,'' in Proceedings ``100th anniversary of the birth of I.Ya. Pomeranchuk", Phys.Usp. 57 (2014) no.2, 152-208
  [arXiv:1312.1214 [hep-th]].

\bibitem{Russo:2014nka} 
  J.~G.~Russo,
  ``$ \mathcal{N} $ = 2 gauge theories and quantum phases,''
  JHEP {\bf 1412}, 169 (2014)
  [arXiv:1411.2602 [hep-th]].



\bibitem{Russo:2018vun}
  J.~G.~Russo and K.~Zarembo,
  ``Wilson loops in antisymmetric representations from localization in supersymmetric gauge theories,''
  Rev.\ Math.\ Phys.\  {\bf 30} (2018) no.07,  1840014
  [arXiv:1712.07186 [hep-th]].

\bibitem{Buchel:2013id}
  A.~Buchel, J.~G.~Russo and K.~Zarembo,
  ``Rigorous Test of Non-conformal Holography: Wilson Loops in N=2* Theory,''
  JHEP {\bf 1303} (2013) 062
  [arXiv:1301.1597 [hep-th]].

\bibitem{Bobev:2013cja}
  N.~Bobev, H.~Elvang, D.~Z.~Freedman and S.~S.~Pufu,
  ``Holography for $N = 2^*$ on $S^4$,''
  JHEP {\bf 1407} (2014) 001
  [arXiv:1311.1508 [hep-th]].
  
\bibitem{Zarembo:2014ooa}
  K.~Zarembo,
  ``Strong-Coupling Phases of Planar N=2* Super-Yang-Mills Theory,''
  Theor.\ Math.\ Phys.\  {\bf 181} (2014) no.3,  1522
  [arXiv:1410.6114 [hep-th]].
  
\bibitem{Chen-Lin:2017pay}
  X.~Chen-Lin, D.~Medina-Rincon and K.~Zarembo,
  ``Quantum String Test of Nonconformal Holography,''
  JHEP {\bf 1704} (2017) 095
  [arXiv:1702.07954 [hep-th]].
  
\bibitem{Bobev:2018hbq}
  N.~Bobev, F.~F.~Gautason and J.~Van Muiden,
  ``Precision Holography for $\mathcal{N}=2^{*}$ on $S^4$ from type IIB Supergravity,''
  JHEP {\bf 1804} (2018) 148
  [arXiv:1802.09539 [hep-th]].
%

\bibitem{Russo:2019lgq} 
  J.~G.~Russo, E.~Widen and K.~Zarembo,
  ``$N$ = 2$^{∗}$ phase transitions and holography,''
  JHEP {\bf 1902}, 196 (2019)
  [arXiv:1901.02835 [hep-th]].
  
  
  
\bibitem{Okuda:2010ke}
  T.~Okuda and V.~Pestun,
  ``On the instantons and the hypermultiplet mass of N=2* super Yang-Mills on $S^{4}$,''
  JHEP {\bf 1203} (2012) 017
  [arXiv:1004.1222 [hep-th]].
  
  
\bibitem{Hollowood:2015oma} 
  T.~J.~Hollowood and S.~P.~Kumar,
  ``Partition function of $ \mathcal{N}={2}^{\ast } $ SYM on a large four-sphere,''
  JHEP {\bf 1512}, 016 (2015)
  [arXiv:1509.00716 [hep-th]].

\bibitem{Brandhuber:1996xk}
  A.~Brandhuber and S.~Stieberger,
  ``Selfdual strings and stability of BPS states in N=2 SU(2) gauge theories,''
  Nucl.\ Phys.\ B {\bf 488} (1997) 199
  [hep-th/9610053].


\bibitem{Bilal:1997st} 
  A.~Bilal and F.~Ferrari,
  ``The BPS spectra and superconformal points in massive N=2 supersymmetric QCD,''
  Nucl.\ Phys.\ B {\bf 516}, 175 (1998)
  [hep-th/9706145].




\bibitem{Argyres:1995jj} 
  P.~C.~Argyres and M.~R.~Douglas,
  ``New phenomena in SU(3) supersymmetric gauge theory,''
  Nucl.\ Phys.\ B {\bf 448}, 93 (1995)
  [hep-th/9505062].
  
  
\bibitem{Argyres:1995xn} 
  P.~C.~Argyres, M.~R.~Plesser, N.~Seiberg and E.~Witten,
  ``New N=2 superconformal field theories in four-dimensions,''
  Nucl.\ Phys.\ B {\bf 461}, 71 (1996)
  [hep-th/9511154].

\bibitem{Eguchi:1996vu} 
  T.~Eguchi, K.~Hori, K.~Ito and S.~K.~Yang,
  ``Study of N=2 superconformal field theories in four-dimensions,''
  Nucl.\ Phys.\ B {\bf 471}, 430 (1996)
  [hep-th/9603002].
  
 

\bibitem{Hanany:1995na}
  A.~Hanany and Y.~Oz,
  ``On the quantum moduli space of vacua of N=2 supersymmetric SU(N(c)) gauge theories,''
  Nucl.\ Phys.\ B {\bf 452} (1995) 283
  [hep-th/9505075].

\bibitem{Argyres:1995wt}
  P.~C.~Argyres, M.~R.~Plesser and A.~D.~Shapere,
  ``The Coulomb phase of N=2 supersymmetric QCD,''
  Phys.\ Rev.\ Lett.\  {\bf 75} (1995) 1699
  [hep-th/9505100].
  
\bibitem{Klemm:1994qs}
  A.~Klemm, W.~Lerche, S.~Yankielowicz and S.~Theisen,
  ``Simple singularities and N=2 supersymmetric Yang-Mills theory,''
  Phys.\ Lett.\ B {\bf 344} (1995) 169
  [hep-th/9411048].
  
\bibitem{Argyres:1994xh}
  P.~C.~Argyres and A.~E.~Faraggi,
  ``The vacuum structure and spectrum of N=2 supersymmetric SU(n) gauge theory,''
  Phys.\ Rev.\ Lett.\  {\bf 74} (1995) 3931
  [hep-th/9411057].

\bibitem{Russo:2015vva} 
  J.~G.~Russo,
  ``Large $N_c$ from Seiberg-Witten Curve and Localization,''
  Phys.\ Lett.\ B {\bf 748}, 19 (2015)
  [arXiv:1504.02958 [hep-th]].


  


\bibitem{Douglas:1995nw} 
  M.~R.~Douglas and S.~H.~Shenker,
  ``Dynamics of SU(N) supersymmetric gauge theory,''
  Nucl.\ Phys.\ B {\bf 447}, 271 (1995)
  [hep-th/9503163].

\bibitem{Ferrari:2001mg} 
  F.~Ferrari,
  ``The Large N limit of N=2 superYang-Mills, fractional instantons and infrared divergences,''
  Nucl.\ Phys.\ B {\bf 612}, 151 (2001)
  [hep-th/0106192].


  
\bibitem{Sonnenschein:1995hv}
  J.~Sonnenschein, S.~Theisen and S.~Yankielowicz,
  ``On the relation between the holomorphic prepotential and the quantum moduli in SUSY gauge theories,''
  Phys.\ Lett.\ B {\bf 367} (1996) 145
  [hep-th/9510129].
  
\bibitem{Eguchi:1995jh}
  T.~Eguchi and S.~K.~Yang,
  ``Prepotentials of N=2 supersymmetric gauge theories and soliton equations,''
  Mod.\ Phys.\ Lett.\ A {\bf 11} (1996) 131
  [hep-th/9510183].

\bibitem{Matone:1995rx}
  M.~Matone,
  ``Instantons and recursion relations in N=2 SUSY gauge theory,''
  Phys.\ Lett.\ B {\bf 357} (1995) 342
  [hep-th/9506102].

\bibitem{DHoker:1996yyu}
  E.~D'Hoker, I.~M.~Krichever and D.~H.~Phong,
  ``The Renormalization group equation in N=2 supersymmetric gauge theories,''
  Nucl.\ Phys.\ B {\bf 494} (1997) 89
  [hep-th/9610156].

\bibitem{Kubota:1997gb}
  T.~Kubota and N.~Yokoi,
  ``Renormalization group flow near the superconformal points in N=2 supersymmetric gauge theories,''
  Prog.\ Theor.\ Phys.\  {\bf 100} (1998) 423
  [hep-th/9712054].

\bibitem{Flume:2004rp}
  R.~Flume, F.~Fucito, J.~F.~Morales and R.~Poghossian,
  ``Matone's relation in the presence of gravitational couplings,''
  JHEP {\bf 0404} (2004) 008
  [hep-th/0403057].

\bibitem{Ohta:1996fr} 
  Y.~Ohta,
  ``Prepotentials of N=2 SU(2) Yang-Mills theories coupled with massive matter multiplets,''
  J.\ Math.\ Phys.\  {\bf 38}, 682 (1997)
  [hep-th/9604059].

\bibitem{D'Hoker:1996nv} 
  E.~D'Hoker, I.~M.~Krichever and D.~H.~Phong,
  ``The Effective prepotential of N=2 supersymmetric SU(N(c)) gauge theories,''
  Nucl.\ Phys.\ B {\bf 489}, 179 (1997)
  [hep-th/9609041].

\bibitem{Masuda:1996xj}
  T.~Masuda and H.~Suzuki,
  ``Periods and prepotential of N=2 SU(2) supersymmetric Yang-Mills theory with massive hypermultiplets,''
  Int.\ J.\ Mod.\ Phys.\ A {\bf 12} (1997) 3413
  [hep-th/9609066].

\bibitem{Masuda:1996np}
  T.~Masuda and H.~Suzuki,
 ``On explicit evaluations around the conformal point in N=2 supersymmetric Yang-Mills theories,''
  Nucl.\ Phys.\ B {\bf 495} (1997) 149
  [hep-th/9612240].


\bibitem{AlvarezGaume:1997fg}
  L.~Alvarez-Gaume, M.~Marino and F.~Zamora,
  ``Softly broken N=2 QCD with massive quark hypermultiplets. 1.,''
  Int.\ J.\ Mod.\ Phys.\ A {\bf 13} (1998) 403
  [hep-th/9703072].

\bibitem{Gaiotto:2010jf}
  D.~Gaiotto, N.~Seiberg and Y.~Tachikawa,
  ``Comments on scaling limits of 4d N=2 theories,''
  JHEP {\bf 1101} (2011) 078
  [arXiv:1011.4568 [hep-th]].
  
\bibitem{Giacomelli:2012ea}
  S.~Giacomelli,
  ``Singular points in N=2 SQCD,''
  JHEP {\bf 1209} (2012) 040
  [arXiv:1207.4037 [hep-th]].

\bibitem{Bolognesi:2015wta}
  S.~Bolognesi, S.~Giacomelli and K.~Konishi,
  ``$ \mathcal{N}=2 $ Argyres-Douglas theories, $ \mathcal{N}=1 $ SQCD and Seiberg duality,''
  JHEP {\bf 1508} (2015) 131
  [arXiv:1505.05801 [hep-th]].
\end{thebibliography}
\end{document}